\documentclass[10pt]{article}
\usepackage{geometry}
\usepackage[dvipdfm]{graphicx}
\usepackage[FIGTOPCAP,nooneline]{subfigure}
\usepackage[T1]{fontenc}
\usepackage{lmodern}
\usepackage{bm}
\usepackage{amsmath, amsfonts, amssymb}
\usepackage{cite}
\usepackage{multirow}
\usepackage{afterpage, array, rotating}
\usepackage{url}
\usepackage{authblk}
\usepackage{color}
\usepackage{algorithm}
\usepackage[noend]{algorithmic}
\usepackage{bm}
\usepackage{mathtools}
\usepackage{ulem}
\usepackage{booktabs}
\usepackage{rotating}
\usepackage{longtable}
\usepackage{pdfpages}
\usepackage{float}

\newcolumntype{A}{>{\centering\arraybackslash}p{5cm}}
\newcolumntype{B}{>{\centering\arraybackslash}p{1.2cm}}
\newcolumntype{C}{>{\centering\arraybackslash}p{3cm}}
\newcolumntype{D}{>{\raggedright\arraybackslash}p{11cm}}
\newcolumntype{E}{>{\raggedleft\arraybackslash}p{1.2cm}}

\title{
Learning Multi-Order Block Structure in Higher-Order Networks
}

\author[1]{Kazuki Nakajima}
\author[2]{Yuya Sasaki}
\author[3]{Takeaki Uno}
\author[1]{Masaki Aida}

\affil[1]{Tokyo Metropolitan University}
\affil[2]{The University of Osaka}
\affil[3]{National Institute of Informatics}

\begin{document}
\date{}
\maketitle

\begin{abstract}
Higher-order networks, naturally described as hypergraphs, are essential for modeling real-world systems involving interactions among three or more entities. 
Stochastic block models offer a principled framework for characterizing mesoscale organization, yet their extension to hypergraphs involves a trade-off between expressive power and computational complexity. 
A recent simplification, a single-order model, mitigates this complexity by assuming a single affinity pattern governs interactions of all orders. 
This universal assumption, however, may overlook order-dependent structural details.
Here, we propose a framework that relaxes this assumption by introducing a multi-order block structure, in which different affinity patterns govern distinct subsets of interaction orders. 
Our framework is based on a multi-order stochastic block model and searches for the optimal partition of the set of interaction orders that maximizes out-of-sample hyperlink prediction performance. 
Analyzing a diverse range of real-world networks, we find that multi-order block structures are prevalent. 
Accounting for them not only yields better predictive performance over the single-order model but also uncovers sharper, more interpretable mesoscale organization.
Our findings reveal that order-dependent mechanisms are a key feature of the mesoscale organization of real-world higher-order networks.
\end{abstract}

\section{Introduction}

Network science has provided a foundational framework for modeling real-world complex systems, representing entities as nodes and their pairwise interactions as edges \cite{boccaletti2006, newman2018}. 
A central goal of network science is to understand the mesoscale organization of complex systems---intermediate-scale patterns that bridge the gap between individual components and the system as a whole \cite{milo2002, girvan2002, colizza2006, clauset2008, newman2018, lotito2022, nakajima2023}. 
A key type of mesoscale organization is community structure: the tendency for networks to be organized into densely connected modules (or communities) that are only sparsely connected to each other \cite{girvan2002, fortunato2016}. 
Uncovering these communities provides profound insights into a system's function, revealing distinct biological modules \cite{ravasz2002, guimera2005}, underlying social groups \cite{zachary1977, lusseau2003}, and constraints on network dynamics \cite{pastorsatorras2015}.

This conventional dyadic approach, however, has fundamental limitations. 
Many real-world phenomena---including conversations in social groups \cite{stehle2011, mastrandrea2015}, co-authorship in scientific collaborations \cite{newman2001, patania2017}, protein interactions in cellular biology \cite{wong2008, gaudelet2018}, and functional brain activity \cite{giusti2016, varley2023}---are inherently higher-order in nature where three or more entities can interact simultaneously. 
Capturing the complex dependencies within these higher-order systems solely through pairwise interactions may obscure their structural properties and lead to inaccurate predictions of their collective dynamics \cite{battiston2020}. 
To address these limitations, complex systems are increasingly modeled as higher-order networks, naturally represented by hypergraphs, spurring the development of a suite of new theories, models, and analytical tools for hypergraphs \cite{battiston2020, battiston2021, boccaletti2023, antelmi2023, majhi2022, lee2025, lotito2023, landry2023}.

Stochastic block models (SBMs) provide a probabilistic, generative framework for extracting community structure from observed networks \cite{airoldi2008, karrer2011, fortunato2016, lee2019, chodrow2021, contisciani2022, sales2023, brusa2024}.
A central challenge in their extension to hypergraphs lies in the trade-off between model expressiveness and computational tractability \cite{contisciani2022, sales2023, ruggeri2023, brusa2024, hood2025}.
This trade-off is exemplified by two opposing approaches. 
On the one hand, a full-order model \cite{sales2023} offers maximum expressiveness by assuming a distinct affinity pattern (i.e., a set of rules for connectivity among communities) for each interaction order (i.e., the number of nodes interacting simultaneously). 
However, this flexibility comes at a high computational complexity, limiting its application to interactions of small orders (e.g., orders 2, 3, and 4) in practice \cite{sales2023}. 
On the other hand, a single-order model \cite{ruggeri2023} ensures scalability by assuming a single, universal affinity pattern governs interactions of all orders. 
While computationally efficient, this strong assumption that the rules of interactions are universal may obscure the diverse organizing principles in complex systems.

Here we address the following question: Do higher-order networks follow a single, universal organizing principle, or are they governed by multiple, order-dependent generative mechanisms? 
By introducing a multi-order SBM for hypergraphs, we explore the optimal partition of interaction orders (i.e., hyperedge sizes) that maximizes out-of-sample hyperlink prediction performance, assuming that interaction orders within the same partition subset share a common affinity pattern.
This allows the model itself to determine whether a single-order or a multi-order model provides a better predictive performance for a given hypergraph.
Our results reveal that multi-order block structure is common across a wide range of empirical networks, including social contact networks and document co-citation networks. 
We demonstrate that accounting for this structure not only improves predictive performance but also uncovers interpretable mesoscale organization of higher-order networks.

\section{Results}

\subsection{Multi-order hypergraph stochastic block model} \label{section:2.1}

We represent a higher-order network as a hypergraph $\mathcal{H} = (\mathcal{V}, \mathcal{E})$, where $\mathcal{V} = \{v_1, \ldots, v_N\}$ is a set of $N$ nodes (i.e., entities) and $\mathcal{E}$ is a set of hyperedges (i.e., interactions) among nodes.
Each hyperedge $e \in \mathcal{E}$ is a subset of $\mathcal{V}$, and its size (i.e., order) $|e|$ is at least two and up to $D$.
We denote by $\Omega$ the set of all possible subsets of $\mathcal{V}$ with sizes ranging from two up to $D$.
We represent the hypergraph by a vector $\mathcal{A} = (A_e)_{e \in \Omega}$, where $A_e$ is a non-negative integer representing the number of times hyperedge $e$ appears in the data \cite{ruggeri2023}.

We employ a statistical inference framework using SBMs for hypergraphs.
While the term ``community'' often implies a densely connected group of nodes (an assortative structure), SBMs provide a more general framework capable of capturing diverse connectivity patterns, including not only assortative but also disassortative or more complex mixing patterns \cite{fortunato2016}. 
We therefore use the term ``community'' to indicate a group of nodes that exhibit similar connectivity profiles.

The single-order model, a mixed-membership SBM for hypergraphs \cite{ruggeri2023}, assumes that latent community structure is defined by two parameter sets \cite{ruggeri2023}:
(i) An $N \times K$ soft membership matrix, $\bm{U}$, where each entry $u_{ik} \geq 0$ represents the propensity of node $v_i \in \mathcal{V}$ belonging to $k$-th community; 
(ii) A symmetric $K \times K$ affinity matrix, $\bm{W}$, where each entry $w_{kq} \geq 0$ governs the density of interactions between $k$-th and $q$-th communities.
This mixed-membership SBM provides an efficient framework for fitting hypergraphs, leveraging computationally inexpensive matrix operations. 
However, this efficiency hinges on the strong, and potentially restrictive, assumption that a single affinity matrix governs interactions of all orders. 
In many real-world systems, interactions of different orders---such as pairwise conversations versus large group meetings among individuals---often play distinct structural and functional roles \cite{battiston2020}. 

We hypothesize that a more descriptive and predictive model can be achieved by allowing different affinity patterns for distinct subsets of interaction orders.
To test this hypothesis, we now extend the single-order model to a multi-order hypergraph SBM, which we refer to as HyperMOSBM.
Let $\mathcal{O} = \{ 2, 3, \ldots, D \}$ be the set of hyperedge sizes ranging from two up to $D$.
We define a partition of these interaction orders, $\mathcal{P} = \{\mathcal{S}_1,\, \mathcal{S}_2,\, \ldots,\, \mathcal{S}_L\}$, into $L$ disjoint subsets, where each subset $\mathcal{S}_l$ groups together interaction orders that are assumed to share a common affinity pattern. 
For example, if $\mathcal{O} = \{2, 3, 4, 5\}$, a possible partition is $\mathcal{P} = \{\{2\}, \{3\}, \{4, 5\}\}$, implying three distinct affinity patterns.
Our multi-order model is parameterized by a collection of $L$ affinity matrices, whereas the single-order model uses a single affinity matrix. 
Specifically, this rule is parameterized by a dedicated symmetric $K \times K$ affinity matrix, $\bm{W}^{(l)}$, where $w_{kq}^{(l)} \geq 0$, for $\mathcal{S}_l$.
The full set of latent parameters in HyperMOSBM is therefore $\bm{\theta} = (\bm{U}, \{\bm{W}^{(l)}\}_{l=1}^L)$.

Based on these definitions, we define the generative model for HyperMOSBM as follows.
We denote by $l_{|e|}$ the index of the set in $\mathcal{P}$ to which the size of hyperedge $e$ belongs (i.e., $|e| \in \mathcal{S}_{l_{|e|}}$).
We assume the weight of hyperedge $e$ follows a Poisson distribution whose rate depends on the subset $\mathcal{S}_{l_{|e|}}$:
\begin{align}
P(A_e\ |\ \bm{U}, \bm{W}^{(l_{|e|})}) &= \exp\left(-\frac{\lambda_e}{\kappa_{|e|}}\right) \frac{\left(\frac{\lambda_e}{\kappa_{|e|}}\right)^{A_e}}{A_e!}.
\label{eq:1}
\end{align}
The definition of $\lambda_e$ mirrors that of the single-order model, with the significant distinction that the affinity matrix $\bm{W}^{(l_{|e|})}$ is selected based on the size of the hyperedge $e$:
\begin{align}
\lambda_e &= \sum_{v_i \in e} \sum_{v_j \in e \backslash \{v_i\}} \sum_{k=1}^K \sum_{q=1}^K u_{ik} \, u_{jq} \, w_{kq}^{(l_{|e|})}. 
\label{eq:2}
\end{align}
The term $\kappa_s = \binom{s}{2} \binom{N-2}{s-2}$ is an order-dependent normalization constant \cite{ruggeri2023}.
The weights of the hyperedges in $\Omega$ are conditionally independent given the full parameter set $\bm{\theta}$:
\begin{align}
P(\mathcal{A}\ |\ \bm{\theta}) = \prod_{e \in \Omega} P(A_e\ |\ \bm{U}, \bm{W}^{(l_{|e|})}).
\label{eq:3}
\end{align}
Consequently, the log-likelihood of the observed hypergraph $\mathcal{A}$ is given by:
\begin{align}
\mathcal{L}(\bm{\theta}\ |\ \mathcal{A}) \triangleq -\sum_{e \in \Omega} \frac{\lambda_e}{\kappa_{|e|}} + \sum_{e \in \mathcal{E}} A_e \log \lambda_e,
\label{eq:4}
\end{align}
where terms independent of $\bm{\theta}$ are discarded. 
The first term involves a sum over all possible hyperedges $\Omega$, which is computationally expensive. 
However, as with the single-order model, this term can be simplified. 
By applying the same reordering technique \cite{ruggeri2023} independently within each subset of sizes $\mathcal{S}_l$, we rearrange this term as:
\begin{align}
    - \sum_{e \in \Omega} \frac{\lambda_e}{\kappa_{|e|}} = - \sum_{l=1}^L C_l \sum_{i=1}^N \sum_{j=1,\ j \neq i}^N \sum_{k=1}^K \sum_{q=1}^K u_{ik} \, u_{jq} \, w_{kq}^{(l)}, \label{eq:5}
\end{align}
where $C_l$ is a constant specific to each subset $\mathcal{S}_l$:
\begin{align}
C_l = \sum_{s \in \mathcal{S}_l} \frac{1}{\kappa_s} \binom{N-2}{s-2}. 
\label{eq:6}
\end{align}
We fit our model to the hypergraph to infer the latent parameters $\bm{\theta}$ by maximizing the evidence lower bound of the log-likelihood (see Section \ref{section:4.1} for details).

A key feature of our model is its flexibility to learn the structure of generative rules from data, by selecting the partition $\mathcal{P}$ of interaction orders.
Rather than selecting a partition that best fits the observed data, which could lead to overfitting, we aim to find the partition that maximizes the model's predictive performance on unseen data. 
To this end, we compute the area under the receiver operating characteristic curve (AUC) in a hyperlink prediction task through a 10-fold cross-validation procedure (see Section \ref{section:4.2} for the details of this procedure). 
The AUC is the probability that a random true positive (an observed hyperedge) is ranked higher than a random true negative (a non-observed hyperedge), with $\text{AUC}=1$ indicating perfect prediction and 0.5 indicating chance. 
The AUC has been adopted as a standard metric for assessing the predictive performance of SBMs \cite{guimera2009, clauset2008, aicher2014, vallescatala2016, ghasemian2020, contisciani2022, ruggeri2023}.

It is worth noting that our framework includes the single-order model as a special case. 
The multi-order model with the trivial partition $\mathcal{P} = \{\mathcal{O}\}$, where all interaction orders are grouped into a single set ($L=1$), is equivalent to the single-order model \cite{ruggeri2023}. 
At the other extreme lies the full-order model, which assumes that each interaction order is governed by an independent affinity matrix (or tensor) \cite{sales2023}, representing the most expressive but computationally intensive approach. 
We define a network as having a multi-order block structure if a non-trivial partition ($L > 1$) achieves a substantially higher AUC score than the single-order model.

\subsection{Validation in synthetic hypergraphs}

\begin{figure*}[p]
\centering
\includegraphics[width=1.0\textwidth]{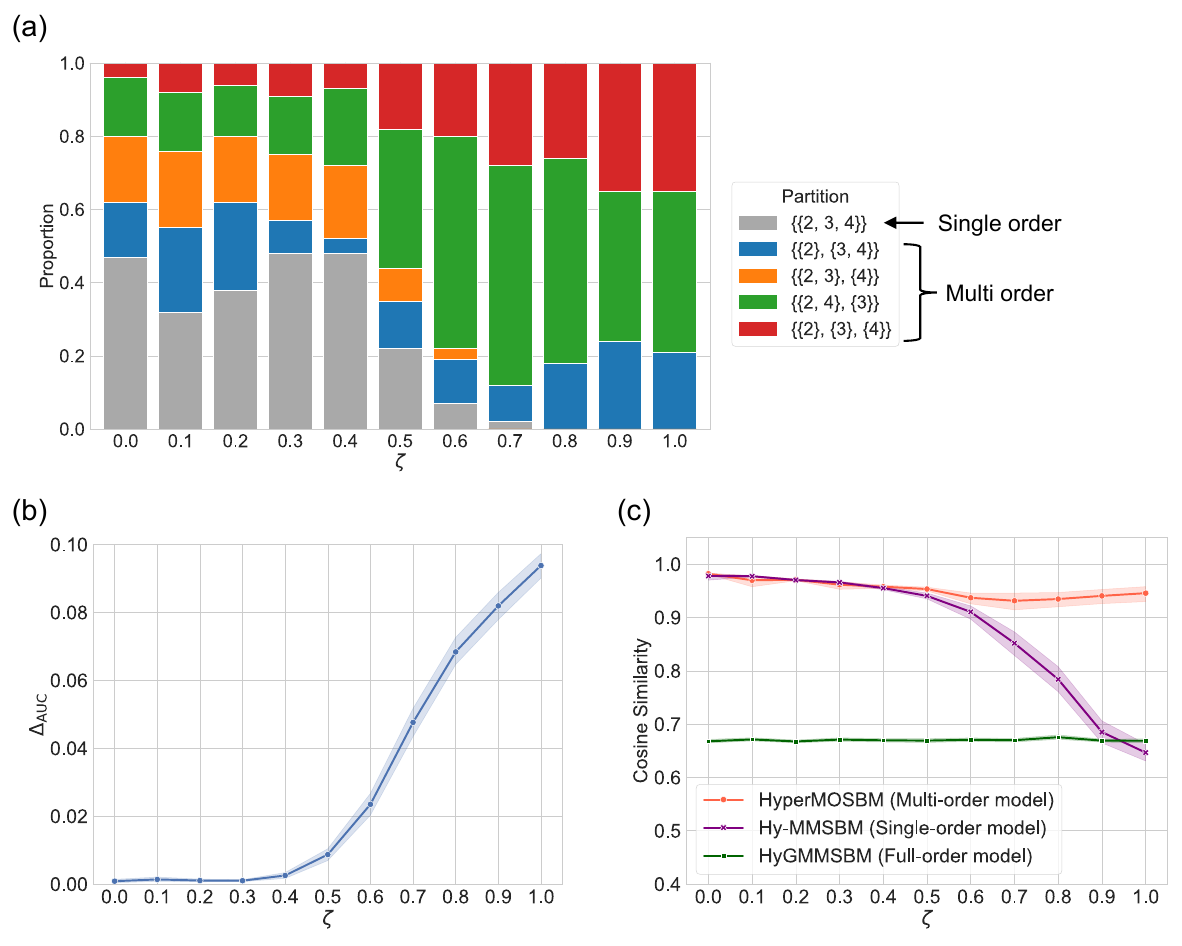}
\caption{
Validation of the multi-order model on synthetic hypergraphs with tunable heterogeneity of affinity patterns across interaction orders. 
All results are averaged over 100 independent instances; shaded regions represent 95\% confidence intervals estimated via bootstrapping.
(a) Proportion of instances where each partition of interaction orders (2, 3, and 4) was selected as optimal by our AUC-driven framework.
(b) Gain in hyperlink prediction performance ($\Delta_{\text{AUC}}$) achieved by the multi-order model with the selected partition compared to the single-order model.
(c) Community recovery performance, measured by cosine similarity, for three models: the multi-order model with the selected partition, the single-order model, and the full-order model.
}
\label{fig:1}
\end{figure*}

We conducted experiments on synthetic hypergraphs with planted ground-truth communities to validate our model.
The objectives here were threefold: first, to verify whether and when our multi-order model yields better hyperlink prediction compared to the single-order model; second, to assess if these gains in predictive performance translate into a more accurate recovery of the ground-truth communities; and third, to benchmark our model's performance in the recovery of ground-truth communities against both the single-order and full-order models.

To these ends, we generated synthetic hypergraphs by adapting a non-uniform hypergraph SBM \cite{ghoshdastidar2017, dumitriu2025}. 
The generated hypergraphs consist of $N=100$ nodes, two equal-size communities, and hyperedges of sizes $s \in \{2, 3, 4\}$. 
The key parameter in our setup is the heterogeneity parameter $\zeta \in [0, 1]$, which controls how the affinity patterns of higher-order interactions ($s \in \{3, 4\}$) deviate from the baseline pairwise pattern ($s=2$). 
We fix the assortative strength of pairwise interactions ($s=2$) at a sufficient level ($a=5$).
This setup allows us to generate a spectrum of hypergraphs, from a uniform single-order block structure at $\zeta=0$ (where the affinity patterns are identical across all hyperedge sizes) to a heterogeneous multi-order block structure at $\zeta=1$ (where the structure becomes disassortative for size $s=3$ and random for size $s=4$).
We generated $100$ independent hypergraphs for each value of $\zeta$ from 0.0 to 1.0 in increments of 0.1. 
The detailed generation procedure is described in Section~\ref{section:4.5}. 

Figure~\ref{fig:1}(a) shows the distribution of optimal partitions selected by our framework for each value of~$\zeta$. 
Contrary to the expectation that the single-order partition would be exclusively selected at $\zeta=0$, it was chosen in only 47\% of instances. 
The remaining instances for $\zeta=0$ favored various multi-order partitions, suggesting that even with a uniform underlying affinity pattern, stochastic fluctuations in the generated hypergraphs may favor the multi-order model.
As $\zeta$ increased from 0.1, the selection of multi-order partitions became more pronounced, although the single-order partition remained a frequent choice (up to 48\%) in the weakly heterogeneous regime ($0.1 \leq \zeta \leq 0.4$). 
However, as $\zeta$ surpassed 0.5, the selection proportion of the single-order partition dropped, and multi-order partitions became dominant.
This trend confirms that our framework selects the underlying multi-order block structure once it becomes a sufficiently prominent feature of the network.

Figure~\ref{fig:1}(b) plots the average gain in predictive performance, $\Delta_{\text{AUC}}$, defined as the AUC of the multi-order model with the partition selected for each instance minus that of the single-order model. 
For $0 \leq \zeta \leq 0.4$, the gain remains marginal. 
However, the performance gain begins to accelerate around $\zeta = 0.5$ and increases steadily thereafter.
This transition to a regime of predictive improvement aligns with the regime where multi-order partitions become dominant (Fig.~\ref{fig:1}(a)).

We then investigated whether the improved predictive performance of our model translates to a more accurate recovery of ground-truth communities.
We compared the performance of our multi-order model (HyperMOSBM) against the single-order (Hy-MMSBM \cite{ruggeri2023}) and the full-order (HyGMMSBM \cite{sales2023}) models. 
We measured performance using the cosine similarity between the inferred and ground-truth community memberships (see Supplementary Section~S1 for the definition) \cite{debacco2017, ruggeri2023}. 

Figure~\ref{fig:1}(c) reveals a clear performance hierarchy among the models. 
While the multi-order model performed comparably to the single-order model in the weakly heterogeneous regime ($0 \leq \zeta \leq 0.5$), it achieved a consistently and substantially higher cosine similarity as heterogeneity increased ($0.5 < \zeta \leq 1.0$). 
In contrast, the full-order model performed poorly across all values of $\zeta$, with its cosine similarity remaining flat and considerably lower than the other two models. 
This counterintuitive finding suggests that the full-order model likely leads to overfitting and that simply maximizing model complexity is not necessarily effective for community recovery in higher-order networks.
Furthermore, the regime where our multi-order model demonstrates its superiority ($0.5 < \zeta \leq 1.0$) aligns with the regime where multi-order partitions are dominantly selected (Fig.~\ref{fig:1}(a)) and where the predictive gain $\Delta_{\text{AUC}}$ becomes substantial (Fig.~\ref{fig:1}(b)).

Our results suggest a practical guideline for model selection. 
While our framework is designed to find the partition with the highest hyperlink prediction performance, a marginal improvement in AUC may not be sufficient to justify abandoning the simpler, single-order model. 
Indeed, for $0 \leq \zeta \leq 0.5$, the marginal or comparable performance in community recovery suggests that the single-order model is preferable due to its simplicity. 
In contrast, for $0.5 < \zeta \leq 1.0$, the improvement in community recovery provides a compelling reason to adopt a multi-order model.
Notably, this transition aligns with a threshold of approximately $\Delta_{\text{AUC}} \ge 0.01$ in our benchmark. 
This heuristic as a threshold for adopting the multi-order model holds largely for other assortative regimes, $a \in \{3, 7, 9\}$ (see Supplementary Section S2).
We therefore propose $\Delta_{\text{AUC}} \ge 0.01$ as a practical heuristic for determining whether a multi-order model is warranted. 

\subsection{Prevalence of multi-order block structure in empirical hypergraphs}

\begin{table*}[t]
\caption{Datasets. $N$: number of nodes. $M$: number of hyperedges. $\bar{k}$: average degree of the node. $\bar{s}$: average size of the hyperedge. $D$: maximum size of the hyperedge. $Z$: number of categories in the node label.}
\label{table:1}
\begin{center}
\begin{tabular}{ l  c  c  c  c  c  c  c  c } \toprule
Data & Category & $N$ & $M$ & $\bar{k}$ & $\bar{s}$ & $D$ & $Z$ & Source \rule[0mm]{0mm}{4.5mm} \\ \midrule
high-school & Contact & 327 & 7,818 & 55.6 & 2.33 & 5 & 9 & \cite{mastrandrea2015, chodrow2021, benson} \\ \addlinespace
primary-school & Contact & 242 & 12,704 & 127 & 2.42 & 5 & 11 & \cite{stehle2011, gemmetto2014, chodrow2021, benson} \\ \addlinespace
hospital-lyon & Contact & 75 & 1,824 & 59 & 2.43 & 5 & 4 & \cite{vanhems2013, landry2023} \\ \addlinespace
invs13 & Contact & 92 & 787 & 17.6 & 2.06 & 4 & 5 & \cite{genois2015, genois2018, landry2023} \\ \addlinespace
invs15 & Contact & 217 & 4,909 & 48.8 & 2.16 & 4 & 12 & \cite{genois2015, genois2018, landry2023} \\ \addlinespace
justice & Co-voting & 38 & 2,826 & 367 & 4.93 & 9 & 2 & \cite{contisciani2022, ruggeri2023} \\ \addlinespace
walmart & Co-purchase & 1,025 & 3,553 & 9.84 & 2.84 & 11 & 10 & \cite{amburg2020, ruggeri2023} \\ \addlinespace
house-committees & Membership & 1,290 & 335 & 9.16 & 35.3 & 81 & 2 & \cite{house_committees, ruggeri2023} \\ \addlinespace
senate-committees & Membership & 282 & 301 & 18.8 & 17.6 & 31 & 2 & \cite{senate_committees, ruggeri2023} \\ \addlinespace
biochem-cocitations & Co-citation & 8,176 & 49,789 & 15.1 & 2.48 & 17 & 14 & \cite{priem2022} and this work \\ \addlinespace
cs-cocitations & Co-citation & 8,618 & 59,161 & 17.5 & 2.55 & 36 & 11 & \cite{priem2022} and this work \\ \addlinespace
math-cocitations & Co-citation & 2,012 & 14,876 & 19.1 & 2.58 & 13 & 10 & \cite{priem2022} and this work \\ \addlinespace
neuro-cocitations & Co-citation & 2,345 & 15,915 & 17.6 & 2.6 & 50 & 8 & \cite{priem2022} and this work \\ \addlinespace
physics-cocitations & Co-citation & 5,390 & 43,942 & 21.8 & 2.68 & 28 & 8 & \cite{priem2022} and this work \\ \addlinespace

\bottomrule
\end{tabular}
\end{center}
\end{table*}

\begin{table*}[p]
\caption{
Comparison of hyperlink prediction performance for 14 empirical hypergraphs. 
The two left columns show the number of subsets $L = |\mathcal{P}_{\text{final}}|$ in the final partition and the mean AUC gain $\Delta_{\text{AUC}}$ over the single-order model. 
Asterisks on $\Delta_{\text{AUC}}$ denote statistical significance based on a one-sided paired $t$-test with Bonferroni correction (*: $p < 0.05/14$, **: $p < 0.01/14$, ***: $p < 0.001/14$).
The three right columns show the mean AUC ($\pm$ standard deviation over 10 splits) for the single-order (Hy-MMSBM), our multi-order (HyperMOSBM), and the full-order (HyGMMSBM) models. 
The model achieving the highest mean AUC for each dataset is indicated in bold. 
For the `invs13' and `justice' datasets, the performance of Hy-MMSBM and HyperMOSBM is identical because our framework selected the single-order partition ($L=1$). 
`DNF' indicates that inference did not complete within the 24-hour time limit.
}
\label{table:2}
\begin{center}
\begin{tabular}{l @{\hspace{3em}} l l @{\hspace{2em}} c c c}
\toprule
 & & & \multicolumn{3}{c}{Hyperlink Prediction AUC} \\
 \cmidrule(l){4-6}
Dataset & $L$ & $\Delta_{\text{AUC}}$ & Hy-MMSBM & HyperMOSBM & HyGMMSBM \\ \midrule
high-school & 2 & 0.015\textsuperscript{***} & 0.918 $\pm$ 0.006 & \textbf{0.933} $\pm$ 0.005 & 0.735 $\pm$ 0.006 \\ \addlinespace
primary-school & 2 & 0.012\textsuperscript{**} & 0.905 $\pm$ 0.007 & \textbf{0.917} $\pm$ 0.005 & DNF \\ \addlinespace
hospital-lyon & 2 & 0.012\textsuperscript{} & 0.871 $\pm$ 0.018 & \textbf{0.884} $\pm$ 0.010 & 0.778 $\pm$ 0.024 \\ \addlinespace
invs13 & 1 & 0.000 & \textbf{0.799} $\pm$ 0.029 & \textbf{0.799} $\pm$ 0.029 & 0.706 $\pm$ 0.022 \\ \addlinespace
invs15 & 2 & 0.009\textsuperscript{} & 0.831 $\pm$ 0.014 & \textbf{0.840} $\pm$ 0.008 & 0.731 $\pm$ 0.007 \\ \addlinespace
justice & 1 & 0.000 & \textbf{0.899} $\pm$ 0.011 & \textbf{0.899} $\pm$ 0.011 & 0.767 $\pm$ 0.026 \\ \addlinespace
walmart & 2 & 0.020\textsuperscript{*} & 0.781 $\pm$ 0.013 & \textbf{0.801} $\pm$ 0.015 & DNF \\ \addlinespace
house-committees & 3 & 0.011\textsuperscript{} & 0.930 $\pm$ 0.019 & \textbf{0.942} $\pm$ 0.022 & DNF \\ \addlinespace
senate-committees & 2 & 0.021\textsuperscript{*} & 0.902 $\pm$ 0.032 & \textbf{0.923} $\pm$ 0.028 & DNF \\ \addlinespace
biochem-cocitations & 3 & 0.024\textsuperscript{***} & 0.886 $\pm$ 0.012 & \textbf{0.911} $\pm$ 0.005 & DNF \\ \addlinespace
cs-cocitations & 2 & 0.016\textsuperscript{***} & 0.935 $\pm$ 0.006 & \textbf{0.952} $\pm$ 0.002 & DNF \\ \addlinespace
math-cocitations & 2 & 0.037\textsuperscript{**} & 0.868 $\pm$ 0.014 & \textbf{0.905} $\pm$ 0.016 & DNF \\ \addlinespace
neuro-cocitations & 3 & 0.030\textsuperscript{**} & 0.839 $\pm$ 0.014 & \textbf{0.869} $\pm$ 0.012 & DNF \\ \addlinespace
physics-cocitations & 3 & 0.026\textsuperscript{**} & 0.905 $\pm$ 0.009 & \textbf{0.931} $\pm$ 0.007 & DNF \\ \addlinespace
\bottomrule
\end{tabular}
\end{center}
\end{table*}

\begin{figure*}[t]
\centering
\includegraphics[width=1.0\textwidth]{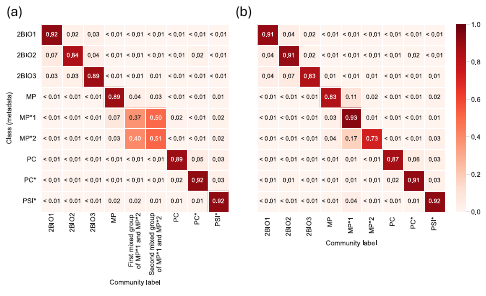}
\caption{
Correspondence between ground-truth classes and inferred communities in the high-school contact network. 
The panels show the averaged membership matrix inferred by (a) the single-order model and (b) our multi-order model.
Each row corresponds to a ground-truth class, and each column represents an inferred community. 
We generated community labels using a large language model based on representative members and their classes (see Methods).
}
\label{fig:2}
\end{figure*}

We now apply our framework to a diverse set of empirical hypergraphs (see Table~\ref{table:1} for summaries and Section~\ref{section:4.6} for details). 
A key challenge in this empirical analysis arises from the combinatorial explosion in the number of possible partitions of the set of interaction orders, $\mathcal{O} = \{2, \ldots, D\}$.
To address this, we adopt a hybrid search strategy: if $|\mathcal{O}| \leq 4$, we perform an exhaustive search to find the globally optimal partition; otherwise, we employ a greedy search algorithm to find a locally optimal partition (see Section~\ref{section:4.3} for full details).

Table~\ref{table:2} summarizes the hyperlink prediction performance for the single-order, full-order, and our multi-order models across the 14 empirical hypergraphs. 
Our framework selected a multi-order partition ($L>1$) in 12 of the 14 datasets, demonstrating the prevalence of multi-order structure. 
In all but one of these cases (`invs15' data), the resulting performance gain $\Delta_{\text{AUC}}$ exceeded the practical threshold of 0.01 we proposed. 
Furthermore, for nine of these datasets, the improvement in AUC was statistically significant after Bonferroni correction (Bonferroni-corrected $p < 0.05/14$). 
Notably, all five co-citation networks were found to possess a significant multi-order block structure.
In contrast, the full-order model was computationally infeasible for most datasets and yielded inferior predictive performance even for those where inference completed.

Beyond the quantitative performance gains, the final partitions of interaction orders qualitatively differ across domains (see Supplementary Table~S2 for the full results). 
For the five contact networks, no single organizing principle emerges; instead, the optimal partitions are diverse.
For example, the partition for the primary-school data separates size-four interactions from the rest ($\{\{2,3,5\}, \{4\}\}$), while the high-school data suggests that size-three interactions behave uniquely ($\{\{2,4,5\}, \{3\}\}$). 
In contrast, in the majority of co-citation networks, the final partition treats pairwise interactions (size two) as distinct from higher-order interactions. 
This may suggest that the mechanism of citing two research papers together differs from that of citing three or more.

The `high-school' hypergraph, one of the contact networks, provides an excellent case to examine how such a structural finding translates to improved descriptive accuracy. 
Its advantage lies in the availability of a known ground-truth community structure (i.e., student classes) \cite{mastrandrea2015, chodrow2021}. 
We visualized an averaged membership matrix for each model by normalizing each student's membership vector to sum to one and then averaging these vectors across all students within the same ground-truth class.
The single-order model exhibits substantial ambiguity, struggling to distinguish between the `MP*1' and `MP*2' classes (Fig.~\ref{fig:2}(a)). 
In contrast, our multi-order model achieves a near one-to-one mapping, largely separating these two classes (Fig.~\ref{fig:2}(b)). 
For this network, our model identified the unique role of size-three interactions by selecting $P_{\text{final}} = \{\{2,4,5\}, \{3\}\}$ as the optimal partition.
Specifically, the inferred affinity patterns suggest that while students in the `MP*1' class exhibit assortative interactions in pairwise conversations, their triadic interactions are relatively sparse (see Supplementary Fig.~S5).
Thus, the improvement in recovering the ground-truth labels stems from our model's flexibility in capturing these order-specific interaction patterns.

\subsection{Case study: Computer science paper co-citation network}

\begin{figure*}[t]
\centering
\includegraphics[width=1.0\textwidth]{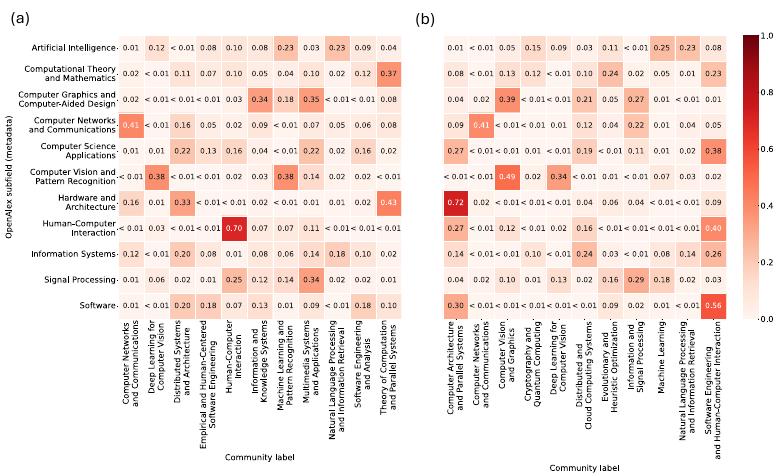}
\caption{
Inferred community structures in the co-citation network.
The panels show the averaged membership matrix inferred by (a) the single-order model and (b) our multi-order model.
Each row corresponds to a computer science subfield, and each column represents an inferred community. 
We assigned community labels based on their representative papers using a large language model (see Methods for details). 
}
\label{fig:3}
\end{figure*}

\begin{figure*}[t]
\centering
\includegraphics[width=1.0\textwidth]{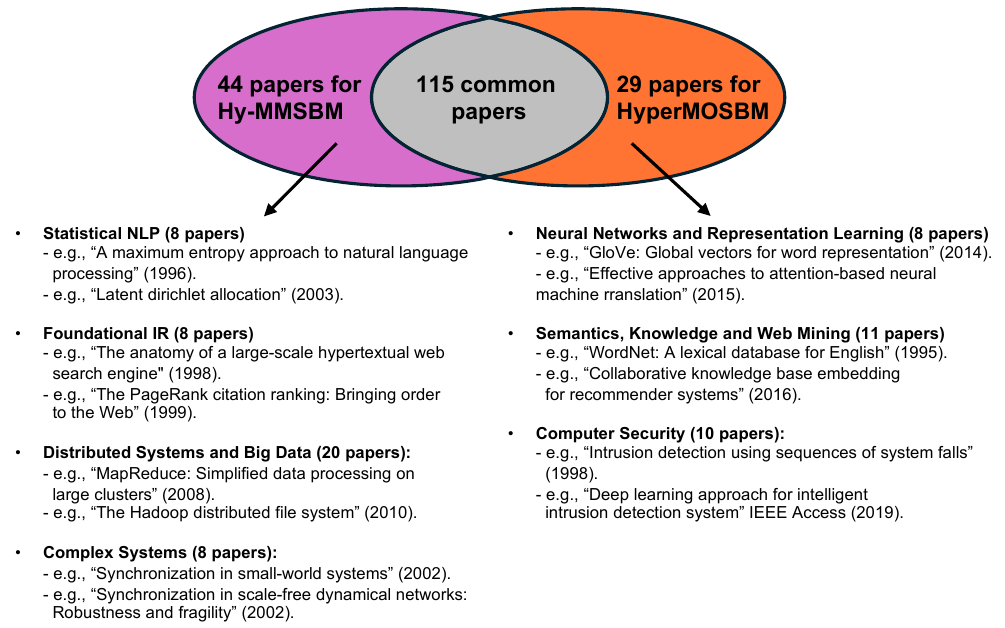}
\caption{
Disentangling the NLP and IR community. 
The Venn diagram compares the sets of papers classified as representative of the NLP and IR community by the single-order (Hy-MMSBM) and our multi-order (HyperMOSBM) models.
}
\label{fig:4}
\end{figure*}

\begin{figure*}[t]
\centering
\includegraphics[width=1.0\textwidth]{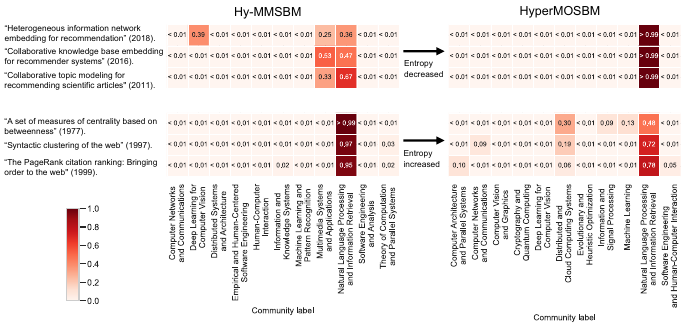}
\caption{
Comparison of community membership vectors for papers with the largest changes in their entropy.
In each panel, the left and right columns correspond to the single-order (Hy-MMSBM) and our multi-order (HyperMOSBM) models, respectively.
Top panel: Top three papers with the largest entropy decrease, selected from those that became representative of the NLP and IR community in the multi-order model.
Bottom panel: Top three papers with the largest entropy increase, selected from those that are no longer classified as representative of the NLP and IR community in the multi-order model. 
}
\label{fig:5}
\end{figure*}

We now focus on the computer science co-citation network, which lacks a predefined ground-truth community structure in contrast to the high-school hypergraph.
In this network, nodes represent highly-cited papers in computer science, and co-citation patterns may reflect the research landscape of the discipline \cite{trujillo2018}. 
This provides an opportunity to assess how each model discovers an interpretable organization from data. 

We compare the averaged membership matrices from both models, which show the correspondence between the inferred communities and the eleven subfields defined by OpenAlex (see Fig.~\ref{fig:3}).
We do not expect either model to produce a one-to-one mapping between the metadata (subfields) and the inferred communities. 
This is because the boundaries between research subfields are themselves inherently blurry and overlapping in the computer science discipline \cite{chakraborty2018}. 
Nevertheless, a careful investigation of the relationship between metadata and inferred communities can reveal the organization captured by each model \cite{peel2017}.
Indeed, a comparison of the two matrices suggests that our multi-order model infers a sharper, interpretable community structure. 
For example, the single-order model generates apparently redundant communities related to software engineering: 'Software Engineering and Analysis', and `Empirical and Human-Centered Software Engineering'. 
In contrast, the multi-order model consolidates these software-related papers into a single, cohesive `Software Engineering and Human-Computer Interaction' community.
Additionally, the single-order model distributes the papers from the `Computer Vision and Pattern Recognition' subfield across three different communities: `Deep Learning for Computer Vision', `Machine Learning and Pattern Recognition', and `Multimedia Systems and Applications'.
Our multi-order model primarily distributes these papers into two communities: `Computer Vision and Graphics' and `Deep Learning for Computer Vision'.

To further investigate these differences at a more granular level, we focus on the `Natural Language Processing and Information Retrieval' (NLP and IR) community. 
Interestingly, both models inferred a community centered on this theme.
This emergent community, which does not correspond directly to a single OpenAlex subfield, warrants deeper investigation, especially given the recent cross-disciplinary impact of large language models \cite{brown2020}. 
We therefore compare the sets of papers that each model classifies as ``representative'' of this community, defined as those with a membership probability for that community of at least 0.9.

Figure~\ref{fig:4} visualizes this comparison as a Venn diagram. 
The 115 core papers, common to both models, comprise topics such as foundational IR, statistical NLP, representation learning for NLP, text mining, resources, and evaluation (see Supplementary Table S3 for paper examples).
The key difference between the models lies in the sets of papers unique to each. 
We manually categorized these papers to understand their thematic composition.
The 44 papers identified exclusively by the single-order model include those on distributed systems and complex systems. 
These papers' direct relevance to core NLP and IR topics is not immediately apparent, and our multi-order model assigns near-zero probabilities to most of them (see Supplementary Table S4).
It is also worth noting that for the `Statistical NLP' and `Foundational IR' papers unique to the single-order model, our multi-order model still assigned membership probabilities of at least 0.7 (see Supplementary Table S4).
Among these 44 papers, 26 (59\%) share no co-citations with the 115 core papers. 
This suggests the single-order model conflates the core topics with generally impactful but thematically distant research.
On the other hand, the set of 29 papers unique to our multi-order model includes those on neural networks and lexical resources, some of which were assigned membership probabilities lower than 0.5 by the single-order model (see Supplementary Table S5).
While this set includes papers from computer security applications, only 12 (41\%) of the 29 papers had no co-citations with the 115 core NLP and IR papers. 
Thus, the papers classified as representative by the multi-order model are, on the whole, more closely related to NLP and IR topics than those classified by the single-order model.

To understand the qualitative differences between the sets of papers unique to each model, we examine individual papers that underwent the most substantial changes in their community assignments. 
We quantify the certainty of community assignment of a paper using the Shannon entropy of its membership vector. 
We focus on two categories: papers that became representative (i.e., their membership probabilities were at least 0.9) of the NLP and IR community in our multi-order model, and those that are no longer classified as such (i.e., their membership probabilities were less than 0.9).

The first category exemplifies the clarification of community assignment (see Fig.~\ref{fig:5}(a)). 
These papers (P1--P3), all of which focus on recommender systems and topic modeling, represent core topics at the intersection of NLP and IR. 
For instance, the papers P1 \cite{shi2019} and P2 \cite{zhang2016} both leverage network embedding to improve the performance of recommender systems.
Similarly, the paper P3 \cite{wang2011} integrates probabilistic topic modeling, a core NLP method, with collaborative filtering for recommending scientific articles. 
Despite their high relevance to NLP and IR, these papers were ambiguously assigned by the single-order model. 
In contrast, our multi-order model decisively assigned them to the NLP and IR community.

The second category illustrates the re-evaluation of interdisciplinary impact (see Fig.~\ref{fig:5}(b)).
These papers (P4--P6), including foundational works on centrality measures and web structure, were narrowly confined to the NLP and IR community by the single-order model, despite their broad impact across computer science and beyond. 
A prime example is the paper on betweenness centrality (P4) \cite{freeman1977}. 
While originally a metric for social network analysis, it is now a fundamental tool across various domains.
Our multi-order model captures this multifaceted nature by distributing the paper's membership across several relevant communities.
Indeed, the paper retains its primary relevance to the NLP and IR community, where centrality measures are crucial for identifying influential authors in citation networks \cite{yan2009} and extracting key terms from text graphs \cite{palshikar2007}.
Simultaneously, it uncovers the paper's strong connection to the `Distributed and Cloud Computing Systems' community, reflecting the extensive research into parallel and distributed algorithms required to compute this metric on large-scale networks \cite{bader2006, you2017}. 
Our model demonstrates a more accurate representation of the interdisciplinary impact of this foundational work.

\section{Discussion}

In this study, we introduced a multi-order SBM to test the hypothesis that real-world higher-order networks possess multi-order block structure, where affinity patterns vary across different subsets of interaction orders. 
Our framework searches for the optimal partition of the set of interaction orders that maximizes out-of-sample hyperlink prediction performance. 
In synthetic hypergraphs, as the heterogeneity of affinity patterns across interaction orders increased, our framework selected a multi-order partition, resulting in accurate recovery of the ground-truth communities. 
Furthermore, the multi-order model provides a better predictive performance and a more interpretable, inferred structure than the single-order model for a majority of the real-world hypergraphs we examined.
Importantly, it also consistently yields better performance in both community recovery and hyperlink prediction compared to the full-order model, which offers the richest representation of affinity patterns but is computationally expensive.
These results provide evidence for the prevalence of multi-order block structure and underscore the significance of accounting for them to build more descriptive and predictive models of higher-order networks.

Our work draws conceptual parallels with the problem of model and structural reducibility in multilayer networks (i.e., pairwise networks where interaction types differ across layers) \cite{dedomenico2015, vallescatala2016, stanley2016}. 
In multilayer networks, a key challenge is to find the most parsimonious layer representation by balancing model fidelity and complexity \cite{kivela2014, boccaletti2014}. 
We argue that a similar trade-off exists in the SBM framework for higher-order networks with respect to interaction order.
At one end lies the single-order model, assuming a single affinity matrix governs interactions of all orders \cite{ruggeri2023}. 
At the other end lies the full-order model, which posits an independent affinity matrix or tensor for interactions of each order \cite{sales2023}. 
While this latter approach is theoretically the most expressive, our empirical results (Table \ref{table:2}) demonstrate that it often suffers from inferior predictive performance and computational limitations.
By extending the computationally efficient, single-order model to accommodate a multi-order partition of interaction orders in its inference procedure, our work provides a principled and practical way to navigate the space between these two extremes. 
We find that, for a majority of the real-world hypergraphs examined, the final partition selected by our framework is a multi-order one (Table \ref{table:2}). 
This suggests that real-world higher-order networks are organized around a finite number of distinct generative mechanisms, each governing a specific range of interaction sizes.
Indeed, as observed in contact and co-citation networks, higher-order interactions can exhibit distinct affinity patterns compared to pairwise interactions.

Higher-order interactions can introduce nontrivial collective phenomena in dynamical processes such as social contagion \cite{iacopini2019, dearruda2020, landry2020, mancastroppa2023}, synchronization \cite{lucas2020, zhang2023}, and cooperation \cite{xu2024, alvarezrodriguez2021}.
While previous studies have often explored the consequences of assuming different dynamical rules for different interaction orders, our framework provides a preceding step: an empirical method to first infer the underlying multi-order block structure from data.
One could then simulate dynamical processes on the hypergraph by assigning distinct dynamical rules to each subset of interaction orders identified by our framework.
Furthermore, our model can produce synthetic hypergraphs that exhibit multi-order block structure. 
By adapting the efficient sampling framework developed for a hypergraph generative model \cite{ruggeri2024}, one can generate an ensemble of such hypergraphs to systematically study how multi-order block structure influences dynamical processes.
Our work can thus contribute to realizing more descriptive and predictive models of dynamical processes by revealing the underlying multi-order block structure of higher-order networks.

We acknowledge several limitations of this study. 
First, our partition selection relies on a supervised criterion, the hyperlink prediction AUC. 
While this metric is a widely adopted standard for assessing the predictive performance of SBMs \cite{guimera2009,  clauset2008, aicher2014, vallescatala2016, ghasemian2020, contisciani2022, ruggeri2023}, it entails a computationally intensive cross-validation procedure. 
A possible alternative is to employ unsupervised criteria, for instance, by quantifying the structural or functional distance between sub-hypergraphs induced by different subsets of interaction orders \cite{surana2023, lucas2024, feng2024, agostinelli2025}. 
Second, the number of candidate partitions of the hyperedge sizes grows superexponentially with the maximum interaction order, posing a scalability challenge. 
Our hybrid search strategy partially mitigates this, but exploring the entire partition space remains computationally prohibitive for networks with large maximum interaction orders. 
One potential approach to partially address this is to first determine an ``effective'' maximum order of the system to reduce the search space. 
Indeed, recent work on dynamical and functional reducibility in higher-order networks has shown that the structural and dynamical contributions of large hyperedges can sometimes be negligible \cite{neuhauser2024, lucas2024}.
Third, we treated the number of latent communities, $K$, as a fixed hyperparameter informed by node metadata in empirical data. 
For example, in the co-citation networks, setting $K$ equal to the number of OpenAlex subfields may be too coarse to capture the inherently fuzzy boundaries between research subfields.
Furthermore, in real-world scenarios where such metadata is unavailable, $K$ must be inferred from the data. 
This might be addressed by inferring $K$ using unsupervised clustering, including modularity clustering \cite{kaminski2019, chodrow2021, kaminski2024}, on the given hypergraph. 
Alternatively, extending a nonparametric Bayesian framework for graphs \cite{latouche2014, peixoto2019} to infer the number of communities in the case of hypergraphs could be fruitful.
Addressing these limitations will be a key step toward a better description and understanding of the structure and dynamics of higher-order networks.

Our work opens up several avenues for future research on higher-order networks.
First, the optimal partition $\mathcal{P}_{\text{final}}$ and its size $L = |\mathcal{P}_{\text{final}}|$ can be treated as new macroscopic features for characterizing higher-order networks. 
Combined with existing hypergraph metrics such as degree correlation \cite{nakajima2022}, hyperedge overlap \cite{lee2021, malizia2025}, and path-based measures \cite{nortier2025}, these metrics could enable a more diverse classification of higher-order systems across diverse domains.
Second, extending our framework to temporal hypergraphs is a promising direction. 
Recent work has begun to uncover complex temporal correlations in higher-order systems \cite{gallo2024}, and our framework could be adapted to track how the roles of different interaction orders evolve over time. 
Finally, integrating node attributes could further enhance predictive accuracy, as combining structure and attributes is a powerful strategy for inference in networks \cite{newman2016, badalyan2024, nakajima2025}. 
Addressing these research directions will be key to building a more comprehensive understanding of the structure and dynamics of real-world higher-order systems.

\section{Methods}

\subsection{Inference of the latent parameters} \label{section:4.1}

We fit the latent parameters $\bm{\theta} = (\bm{U}, \{\bm{W}^{(l)}\}_{l=1}^L)$ to the observed hypergraph $\mathcal{A}$ by maximizing the log-likelihood function given in Eq.~\eqref{eq:4}. Since a direct analytical maximization of $\mathcal{L}(\bm{\theta}\ |\ \mathcal{A})$ is intractable, we employ an expectation-maximization (EM) algorithm \cite{dempster1977}.
The core of the EM approach is to maximize a tractable lower bound of the log-likelihood. 
We derive this bound by applying Jensen's inequality to the second term of Eq.~\eqref{eq:4}:
\begin{align}
\sum_{e \in \mathcal{E}} A_e \log \lambda_e
&\geq \sum_{e \in \mathcal{E}} A_e \sum_{v_i \in e} \sum_{v_j \in e \backslash \{v_i\}} \sum_{k=1}^K \sum_{q=1}^K \rho_{ijkq}^{(e)} \log \left( \frac{u_{ik} \, u_{jq} \, w_{kq}^{(l_{|e|})}}{\rho_{ijkq}^{(e)}} \right) 
\label{eq:7}
\end{align}
where $\rho_{ijkq}^{(e)}$ is an arbitrary variational probability distribution satisfying $\sum_{v_i \in e} \sum_{v_j \in e \backslash \{v_i\}} \sum_{k,q} \rho_{ijkq}^{(e)} = 1$ for each $e \in \mathcal{E}$. 
The inequality becomes an equality (i.e., the bound is tightest) when we set the variational distribution as the posterior probability of the latent variables:
\begin{align}
\rho_{ijkq}^{(e)} = \frac{u_{ik} \, u_{jq} \, w_{kq}^{(l_{|e|})}}{\lambda_e}.
\label{eq:8}
\end{align}
By substituting this lower bound into the log-likelihood function, we obtain the objective function for the EM algorithm, often called the evidence lower bound (ELBO):
\begin{align}
F(\rho, \bm{\theta}\ |\ \mathcal{A}) 
\triangleq &- \sum_{l=1}^L C_l \sum_{i=1}^N \sum_{j=1,\ j \neq i}^N \sum_{k=1}^K \sum_{q=1}^K u_{ik} \, u_{jq} \, w_{kq}^{(l)} \notag \\
&+ \sum_{e \in \mathcal{E}} A_e \sum_{v_i \in e} \sum_{v_j \in e \backslash \{v_i\}} \sum_{k=1}^K \sum_{q=1}^K \rho_{ijkq}^{(e)} \log \left( \frac{u_{ik} \, u_{jq} \, w_{kq}^{(l_{|e|})}}{\rho_{ijkq}^{(e)}} \right).
\label{eq:9}
\end{align}
We then maximize this ELBO with respect to $\bm{\rho}$ and $\bm{\theta}$ by alternating between two steps:
\begin{itemize}
    \item E-step: With $\bm{\theta}$ fixed, we maximize the ELBO with respect to the variational distribution $\rho$ by updating it according to Eq.~\eqref{eq:8}.
    \item M-step: With $\rho$ fixed, we maximize the ELBO with respect to the latent parameters $\bm{\theta}$.
\end{itemize}
The detailed update rules for the M-step are derived as follows.

We first derive the update rule for the membership matrix $\bm{U}$. 
We do not impose a summation-to-one constraint on the membership vectors $\bm{u}_i$, as in Ref.~\cite{ruggeri2023}. 
Therefore, we can find the update for each $u_{ik}$ by taking the partial derivative of the ELBO, $F(\bm{\rho}, \bm{\theta}\ |\ \mathcal{A})$, with respect to $u_{ik}$ and setting it to zero.
The partial derivative is given by:
\begin{align}
\frac{\partial}{\partial u_{ik}} F(\rho, \bm{\theta}\ |\ \mathcal{A}) = \frac{1}{u_{ik}} \left(\sum_{e \in \mathcal{E},\ v_i \in e} A_e \sum_{v_j \in e \backslash \{v_i\}} \sum_{q=1}^K \rho_{ijkq}^{(e)} \right) - \sum_{l=1}^L C_l \sum_{j=1,\ j \neq i}^N \sum_{q=1}^K u_{jq} \, w_{kq}^{(l)}.
\label{eq:10}
\end{align}
Setting the derivative to zero and solving for $u_{ik}$ yields the multiplicative update rule:
\begin{align}
u_{ik} = \frac{\sum_{e \in \mathcal{E},\ v_i \in e} A_e \sum_{v_j \in e \backslash \{v_i\}} \sum_{q=1}^K \rho_{ijkq}^{(e)}}{\sum_{l=1}^L C_l \sum_{j=1,\ j \neq i}^N \sum_{q=1}^K u_{jq} \, w_{kq}^{(l)}}.
\label{eq:11}
\end{align}

Next, we focus on updating $w_{kq}^{(l)}$.
The partial derivative of $F(\rho, \bm{\theta}\ |\ \mathcal{A})$ with respect to $w_{kq}^{(l)}$ is given by
\begin{align}
\frac{\partial}{\partial w_{kq}^{(l)}} F(\rho, \bm{\theta}\ |\ \mathcal{A})
= \frac{1}{w_{kq}^{(l)}} \left( \sum_{e \in \mathcal{E},\ |e| \in \mathcal{S}_l} A_e \sum_{v_i \in e} \sum_{v_j \in e \backslash \{v_i\}} \rho_{ijkq}^{(e)} \right) - C_l \sum_{i=1}^N \sum_{j=1,\ j \neq i}^N u_{ik} \, u_{jq}.
\label{eq:12}
\end{align}
Setting the partial derivative of $F(\rho, \bm{\theta}\ |\ \mathcal{A})$ with respect to $w_{kq}^{(l)}$ to zero yields
\begin{align}
w_{kq}^{(l)} = \frac{\sum_{e \in \mathcal{E},\ |e| \in \mathcal{S}_l} A_e \sum_{v_i \in e} \sum_{v_j \in e \backslash \{v_i\}} \rho_{ijkq}^{(e)}}{C_l \sum_{i=1}^N \sum_{j=1,\ j \neq i}^N u_{ik} \, u_{jq}}.
\label{eq:13}
\end{align}

A naive implementation of these updates would be computationally prohibitive. 
However, by reformulating the updates for our multi-order model based on the efficient matrix-based formulation for the single-order model in \cite{ruggeri2023}, the updates in our model remain efficient. 
The computational complexity per EM iteration is $O(L N K^2 + E K^2 + M K)$, where $E = |\mathcal{E}|$ is the number of hyperedges and $M = \sum_{e \in \mathcal{E}} |e|$ is the sum of the hyperedges' sizes.

In all experiments, we treat the number of latent communities, $K$, as a fixed hyperparameter. 
For the analyses on synthetic hypergraphs, we set $K$ to the number of ground-truth communities. 
For the empirical hypergraphs, where the ground truth is unknown, we set $K$ to the number of unique categories present in the node metadata.
We apply the same value of $K$ to all three models to ensure a consistent comparison.
Further implementation details are provided in Supplementary Section S4.

\subsection{Partition selection} \label{section:4.2}

A central challenge in our multi-order model is to determine the optimal partition of the set of hyperedge sizes, $\mathcal{P}$.  
To address this partition selection problem, we employ a hyperlink prediction task \cite{libennowell2007, chen2024}.
The goal is to select the partition $\mathcal{P}$ that maximizes the model's ability to predict held-out hyperedges when trained on a partial dataset.

To evaluate each candidate partition, we employ a 10-fold cross-validation scheme. 
We partition the set of all observed hyperedges, $\mathcal{E}$, into 10 disjoint folds. 
In each iteration, one fold is treated as the test set, $\mathcal{E}_{\text{test}}$, while the remaining nine folds constitute the training set, $\mathcal{E}_{\text{train}}$ (here, 10\% and 90\% of $\mathcal{E}$, respectively). 
From these sets, we construct the training adjacency vector $\mathcal{A}_{\text{train}}$ and the test adjacency vector $\mathcal{A}_{\text{test}}$. 
Specifically, the entry for a potential hyperedge $e \in \Omega$ in the vector $\mathcal{A}_{\text{train}}$ is the observed weight $A_e$ if $e \in \mathcal{E}_{\text{train}}$, and 0 otherwise. 
The vector $\mathcal{A}_{\text{test}}$ is defined analogously.

We infer the latent parameters by fitting the model to $\mathcal{A}_{\text{train}}$ exclusively.
Given the inferred parameters $\hat{\bm{\theta}}$, we define a scoring function $f_{\hat{\bm{\theta}}}: \Omega \to \mathbb{R}$ for a potential hyperedge $e \in \Omega$ using a quantity monotonically related to its existence probability. 
Under the Poisson assumption (Eq.~\eqref{eq:1}), this probability is given by $P(A_e>0) = 1 - \exp(-\lambda_e/\kappa_{|e|})$. 
Since this probability is monotonically increasing with respect to the rate parameter $\lambda_e / \kappa_{|e|}$, we define our scoring function as $f_{\hat{\bm{\theta}}}(e) = \log\hat{\lambda}_e - \log\kappa_e$, where $\hat{\lambda}_e$ is calculated with the inferred parameters $\hat{\bm{\theta}}$.

We then assess how well the trained model can distinguish between a set of observed hyperedges (true positives) and a set of non-observed hyperedges (true negatives) based on a score assigned to each potential hyperedge.
A true positive is a hyperedge $e \in \mathcal{E}_{\text{test}}$, while a true negative is a potential hyperedge $e' \in \Omega \backslash \mathcal{E}$ that was not present in the original hypergraph.
The AUC is mathematically equivalent to the probability that a randomly chosen true positive is assigned a higher score than a randomly chosen true negative. 
To estimate this value, we perform a pairwise comparison via Monte Carlo sampling. 
We first construct a set of $10^4$ positive-negative pairs, $\mathcal{R} = \{(e_i, e'_i)\}_{i=1}^{10^4}$, where this number of samples is chosen to ensure that the AUC is estimated with a precision of approximately $\pm 0.01$ \cite{ghasemian2020}.
Each positive instance $e_i$ is sampled uniformly at random from $\mathcal{E}_{\text{test}}$ with replacement. 
To construct a balanced set of negative instances, each corresponding negative instance $e'_i$ is sampled from the set of all non-observed hyperedges ($\Omega \backslash \mathcal{E}$) such that its size, $|e'_i|$, matches the size of its positive counterpart, $|e_i|$ \cite{ruggeri2023}. 
The AUC is then estimated as the fraction of pairs in which the positive instance is ranked higher than the negative one:
\begin{align}
\text{AUC} = \frac{1}{10^4} \sum_{i=1}^{10^4} \left[ \mathbb{I}(f_{\hat{\bm{\theta}}}(e_i) > f_{\hat{\bm{\theta}}}(e'_i)) + 0.5 \cdot \mathbb{I}(f_{\hat{\bm{\theta}}}(e_i) = f_{\hat{\bm{\theta}}}(e'_i)) \right],
\label{eq:14}
\end{align}
where $\mathbb{I}(\cdot)$ is the indicator function, and ties are handled by adding 0.5.

This procedure is repeated for all 10 folds. 
The final performance score for a given partition, $\text{AUC}(\mathcal{P})$, is then taken as the mean of the AUC values calculated across the 10 folds.

\subsection{Search algorithm for the partition $\mathcal{P}$} \label{section:4.3}

Finding the partition of the set of unique hyperedge sizes that maximizes the predictive performance is a computationally demanding task. 
The total number of possible partitions of a set with $|\mathcal{O}|$ elements is given by the $|\mathcal{O}|$-th Bell number \cite{graham1989}, which grows superexponentially. 
Given this combinatorial explosion, an exhaustive search is not feasible for many real-world hypergraphs. 
Therefore, we adopt a hybrid strategy: if $|\mathcal{O}| \leq 4$ (i.e., the number of possible partitions is at most 15), we perform an exhaustive search to find the globally optimal partition. 
Otherwise ($|\mathcal{O}| > 4$), we employ a greedy search algorithm to find a locally optimal partition, as described below.

To prevent partitions from being determined by an insufficient number of observed hyperedges (i.e., fitting to sparse data), we only evaluate candidate partitions for which each subset of hyperedge sizes yields a sufficient number of hyperedges. 
Specifically, we require the number of hyperedges whose sizes belong to any subset $\mathcal{S}_l$ to be at least $c$ times the number of parameters in the corresponding affinity matrix, i.e., $\sum_{s \in \mathcal{S}_l} m_s \geq c K (K+1)/2$, where $m_s$ denotes the number of hyperedges of size $s$, and we set $c = 5$. 
This threshold is applied to both the exhaustive and greedy search procedures.

The greedy search algorithm is an iterative procedure that begins with the single-order partition, $\mathcal{P}_1 = \{\mathcal{O}\}$, and attempts to find a better partition in a step-by-step manner. 
In each step, the algorithm refines the current partition, $\mathcal{P}_i$, by generating a set of candidate partitions. 
These candidates are formed by taking every subset $\mathcal{S}_l \in \mathcal{P}_i$ with $|\mathcal{S}_l| \geq 2$ and creating all possible binary splits between adjacent, sorted sizes. 
For example, a set $\{s_1, s_2, s_3, s_4\}$ where $s_1 < s_2 < s_3 < s_4$ would yield three candidate splits: $(\{s_1\}, \{s_2, s_3, s_4\})$, $(\{s_1, s_2\}, \{s_3, s_4\})$, and $(\{s_1, s_2, s_3\}, \{s_4\})$.
For each of these candidate partitions that satisfies the above criterion, we calculate its mean AUC score. 
The algorithm then identifies the single split that results in the partition with the highest AUC, let this be $\mathcal{P}_{\text{cand}}$. 
If $\text{AUC}(\mathcal{P}_{\text{cand}}) - \text{AUC}(\mathcal{P}_i) > 10^{-3}$, the search proceeds to the next step with $\mathcal{P}_{i+1} = \mathcal{P}_{\text{cand}}$. 
Otherwise, or if all subsets have become singletons (i.e., $|\mathcal{S}_l|=1$ for all $l = 1, \ldots, L$), the iteration terminates. 
The final partition, $\mathcal{P}_{\text{final}}$, is used for our analysis, and the performance gain is defined as $\Delta_{\text{AUC}} = \text{AUC}(\mathcal{P}_{\text{final}}) - \text{AUC}(\mathcal{P}_1)$.

Our model, following the original single-order model \cite{ruggeri2023}, assumes that the set of possible interaction orders $\mathcal{O}$ is a consecutive range of integers from 2 to $D$.
Therefore, it is important to note that a resulting partition $\mathcal{P}_{\text{final}}$ may contain subsets of sizes that were not actually present in the empirical data. 
This reflects that the partition describes the structure of the underlying generative process, not just the specific orders realized in a finite sample of hyperedges.

\subsection{Synthetic hypergraphs} \label{section:4.5}

To test our multi-order model, we employ a modified version of the non-uniform hypergraph SBM \cite{ghoshdastidar2017, dumitriu2025}. 
We generate synthetic hypergraphs with $N$ nodes and two ground-truth communities of equal size ($N / 2$). 
Nodes are assigned to one of the two communities with a hard membership.

We generate hypergraphs composed of hyperedges of sizes $s \in \{ 2 , 3 , 4 \}$. 
The existence of each potential hyperedge $e \subset \mathcal{V}$ a given size $s$ is determined independently by a Bernoulli trial. 
The probability of its formation is $p_s$ if all its constituent nodes belong to the same community, and $q_s$ otherwise.
We define these probabilities as $p_s = \tau a_s / \binom{N-1}{s-1}$ and $q_s = \tau b_s / \binom{N-1}{s-1}$ for $s \in \{2, 3, 4\}$, following \cite{dumitriu2025}, where $\tau$ is a global scaling factor that adjusts the expected number of hyperedges to which a node belongs to a target value, $\bar{k}$.

We set the parameters, $a_s$ and $b_s$, for each size $s \in \{2, 3, 4\}$ as follows.
For pairwise interactions ($s=2$), we set $(a_2, b_2) = (a, b)$.
The pairwise interaction pattern is defined as assortative when $a > b$.
For higher-order interactions ($s \in \{3, 4\}$), we introduce a parameter $\zeta \in [0, 1]$ to control their affinity patterns relative to the pairwise pattern:
\begin{align}
a_3 &= (1 - \zeta) a_2 + \zeta b_2 \label{eq:15} \\
b_3 &= (1 - \zeta) b_2 + \zeta a_2 \label{eq:16} \\
a_4 &= (1 - \zeta) a_2 + \frac{\zeta}{2} (a_2 + b_2) \label{eq:17} \\
b_4 &= (1 - \zeta) b_2 + \frac{\zeta}{2} (a_2 + b_2) \label{eq:18}
\end{align}
When $\zeta=0$, the higher-order interactions ($s \in \{3, 4\}$) exhibit the same pattern as the pairwise ones ($s=2$).
As $\zeta$ increases, the interaction pattern for size $s=3$ transitions towards a disassortative pattern (i.e., $a_3 < b_3$ for $\zeta > 0.5$), while the pattern for size $s=4$ converges to a random-like pattern (i.e., $a_4 \approx b_4$).
In our experiments, we set $N = 100$ and $\bar{k} = 20$, and we vary $a \in \{1, 3, 5, 7, 9\}$ while fixing $b=1$.

\subsection{Empirical hypergraphs} \label{section:4.6}

We applied our framework to 14 empirical hypergraphs obtained from multiple data sources, summarized in Table~\ref{table:1}.
To ensure a consistent format across these sources, we treat these hypergraphs as unweighted hypergraphs (i.e., for each potential hyperedge $e \in \Omega$, $A_e = 1$ if $e$ appears in the data, and $A_e = 0$ otherwise).

We use five social contact hypergraphs. 
The `high-school' hypergraph is a social contact network, where nodes are students in a high school, a hyperedge is a face-to-face contact event among a set of students, and each node's label represents the class to which the student belonged \cite{mastrandrea2015}. 
The `primary-school' hypergraph is also a social contact network, where nodes are individuals (i.e., students or teachers), a hyperedge represents an event in which a set of individuals are in face-to-face contact with each other, and each node's label indicates whether the individual is a teacher or to which class a student belongs \cite{stehle2011, gemmetto2014}. 
The `hospital-lyon' hypergraph is a social contact network, where nodes are individuals (i.e., patients or healthcare workers) in a hospital, a hyperedge is a face-to-face contact event among a set of individuals, and each node's label represents their role (i.e., paramedical staff, patient, medical doctor, or administrative staff) \cite{vanhems2013}. 
The `invs13' and `invs15' hypergraphs are social contact networks of individuals in an office building, a hyperedge represents an event in which a set of individuals are in face-to-face contact with each other, and each node's label represents the department to which the individual belonged, with data collected in two different years \cite{genois2015, genois2018}.
We collected the high-school and primary-school hypergraphs from Ref.~\cite{benson} and the hospital-lyon, invs13, and invs15 hypergraphs from Ref.~\cite{landry2023}. 

The `house-committees' and `senate-committees' hypergraphs capture membership structures in the U.S. Congress \cite{house_committees, senate_committees}. 
In these hypergraphs, nodes represent legislators, and a hyperedge consists of all members belonging to a specific committee. 
Each node's label indicates the political party of the legislator (i.e., Democrat or Republican). 
The `justice' hypergraph represents the co-voting patterns of Justices of the U.S. Supreme Court from 1946 to 2019, where nodes are Justices, and a hyperedge connects all Justices who cast the same vote in a particular case \cite{contisciani2022}. 
Each node's label represents the political party affiliation of the Justice. 
The `walmart' hypergraph describes co-purchase behavior, where nodes are products and a hyperedge comprises a set of products bought together in a single transaction \cite{amburg2020}. 
Each node's label indicates the category of the product. 
We collected these four hypergraphs from Ref.~\cite{ruggeri2023}.

Finally, to create a new benchmark for evaluating our model, we constructed five new co-citation hypergraphs for the following research fields: Mathematics (`math-cocitations'), Computer Science (`cs-cocitations'), Biochemistry, Genetics and Molecular Biology (`biochem-cocitations'), Physics and Astronomy (`physics-cocitations'), and Neuroscience (`neuro-cocitations'). 
In these hypergraphs, modular structure emerging from co-citation patterns can reflect the topical organization of research subfields \cite{trujillo2018}.
We followed a consistent procedure to construct each hypergraph using the OpenAlex data (September 2024 snapshot \cite{priem2022}). 
First, for each of the five fields, we defined the set of nodes by selecting `highly-cited papers' from each of its constituent subfields. 
Specifically, for the papers published between 1950 and 2024, we identified the minimal set that received the highest number of citations within each subfield and accounted for at least 10\% of the total citations in that subfield.
We ensured at least 100 papers per subfield. 
The union of these papers formed the node set, with each node labeled by its own subfield.
Second, we formed hyperedges based on co-citation patterns within each field. 
For every paper published between 1950 and 2024 within a given field, we identified the subset of highly-cited papers from the same field that it cited. 
If this subset contained two or more highly-cited papers, it was treated as a hyperedge candidate. 
To filter out incidental co-citations, we retained only those candidates that appeared at least three times. 
Finally, the nodes in each hypergraph were restricted to only those highly-cited papers that belonged to at least one such retained hyperedge.

\subsection{Community labeling using a large language model}

To interpret the community membership matrices inferred by the single-order and multi-order models for the `high-school' and `cs-cocitations' networks, we employed a large language model (LLM; we used Gemini 2.5 Pro in November 2025) to generate descriptive labels for the communities.  
We applied the exact same procedure and prompt template to the results of both models to ensure a fair comparison. 
First, for each of the $K$ inferred communities, we identified the representative nodes, defined as those with a membership probability of at least 0.9 for that community. 
Second, we constructed a prompt containing the metadata and the membership probabilities of these representative nodes. 

For the `high-school' network, the prompt included the class labels of the students along with their membership probabilities. 
We instructed the LLM to act as an expert in social network analysis to analyze the class composition of the representative members and generate a label summarizing the social group (e.g., a specific class or a mixture of classes). 
To facilitate a comparison between the inferred communities and the ground-truth metadata, we required the LLM to adopt the original class name as the label if the community dominantly represented that class.

For the `cs-cocitations' network, the prompt included the titles, publication years, and membership probabilities of the papers. 
We instructed the LLM to act as an expert in bibliographic analysis to first extract 5 to 10 keywords from the titles and then assign a label consistent with the ACM Computing Classification System \cite{acm_ccs} based on these keywords and the paper list. 
As in the high-school network, we instructed the LLM to prioritize the original subfield names from the OpenAlex metadata if they accurately described the community.

While this approach provides an automated, practical means to translate the inferred communities into human-readable descriptions \cite{edge2024}, we acknowledge potential limitations. 
The labels generated by LLMs are probabilistic summaries derived from their training data. 
They may not fully capture the implicit social organization within the high school or the precise, evolving definitions of specialized research subfields in computer science as perceived by actual domain experts.

\section*{Acknowledgments}

We thank Dr.~Maxime Lucas for helpful discussions.
K.N.~thanks the financial support by the JST ACT-X Grant Number JPMJAX24CI and the Nakajima Foundation.
K.N.~and Y.S.~thanks the financial support by the JST ASPIRE Grant Number JPMJAP2328.
K.N.~and M.A.~thanks the financial support by the JSPS KAKENHI Grant Number JP25H01122.

\section*{Declaration of Competing Interest}
The authors declare no competing interests.

\section*{Data Availability}
The `high-school' and `primary-school' networks were obtained from Ref.~\cite{benson}.
The `hospital-lyon', `invs13', and `invs15' hypergraphs from Ref.~\cite{landry2023}.  
The `house-committees', `senate-committees', `justice', and `walmart' are collected from Ref.~\cite{ruggeri2023}.
The five co-citation networks were newly constructed for this study using OpenAlex data (September 2024 snapshot \cite{priem2022}).
The hypergraph data used to generate all figures and tables in this study are available at \url{https://doi.org/10.5281/zenodo.17713331}.

\section*{Code Availability}
The code used for this study is publicly available at \url{https://doi.org/10.5281/zenodo.17713331}.
The single-order model (Hy-MMSBM) was used via the hypergraphx library (version 1.7.8) \cite{lotito2023}.
The full-order model (HyGMMSBM) was obtained from the authors' original repository at \url{https://github.com/seeslab/HyGMMSBM}.

\newpage

\begin{center}
\vspace*{12pt}
{\Large Supplementary Information for:\\
\vspace{12pt}
Learning Multi-Order Block Structure in Higher-Order Networks}
\vspace{12pt} \\
\end{center}

\setcounter{figure}{0}
\setcounter{table}{0}
\setcounter{section}{0}

\renewcommand{\thesection}{S\arabic{section}}
\renewcommand{\thefigure}{S\arabic{figure}}
\renewcommand{\thetable}{S\arabic{table}}
\renewcommand{\theequation}{S\arabic{equation}}
\renewcommand{\thealgorithm}{S\arabic{algorithm}}

\begin{center}
\author{Kazuki Nakajima, Yuya Sasaki, Takeaki Uno, and Masaki Aida}
\vspace{24pt} \\
\end{center}

\section{Evaluation metric for recovery of ground-truth communities}

In the experiments on synthetic hypergraphs, we assessed the performance in recovering the ground-truth communities by comparing the inferred soft membership matrix, $\hat{\bm{U}}$, with the ground-truth hard membership matrix, $\bm{U}$. 
For this purpose, we adopted the permutation-invariant cosine similarity (CS), following Refs.~\cite{debacco2017, ruggeri2023}. 
This metric addresses the label switching problem by finding the optimal permutation of community labels that maximizes the average similarity.
The metric is defined as:
\begin{align}
\text{CS}(\bm{U}, \hat{\bm{U}}) = \max_{\pi \in \Pi_K} \frac{1}{N} \sum_{i=1}^{N} \frac{\bm{u}_i \cdot (\hat{\bm{u}}_i \circ \pi)}{\|\bm{u}_i\|_2 \|\hat{\bm{u}}_i\|_2},
\label{eq:s1}
\end{align}
where:
\begin{itemize}
    \item $\bm{u}_i$ and $\hat{\bm{u}}_i$ are the $K$-dimensional row vectors representing the ground-truth and inferred community memberships for node $i$, respectively.
    \item $\Pi_K$ is the set of all $K!$ permutations of the community indices $\{1, \ldots, K\}$.
    \item $\hat{\bm{u}}_i \circ \pi$ denotes the vector obtained by permuting the elements of $\hat{\bm{u}}_i$ according to a permutation $\pi \in \Pi_K$.
    \item $\cdot$ denotes the dot product, and $\|\cdot\|_2$ denotes the Euclidean norm.
\end{itemize}
The resulting score ranges from 0 to 1, with 1 indicating a perfect match up to a permutation of the labels.

\section{Additional results for synthetic hypergraphs}

\begin{figure*}[p]
\centering
\includegraphics[width=1.0\textwidth]{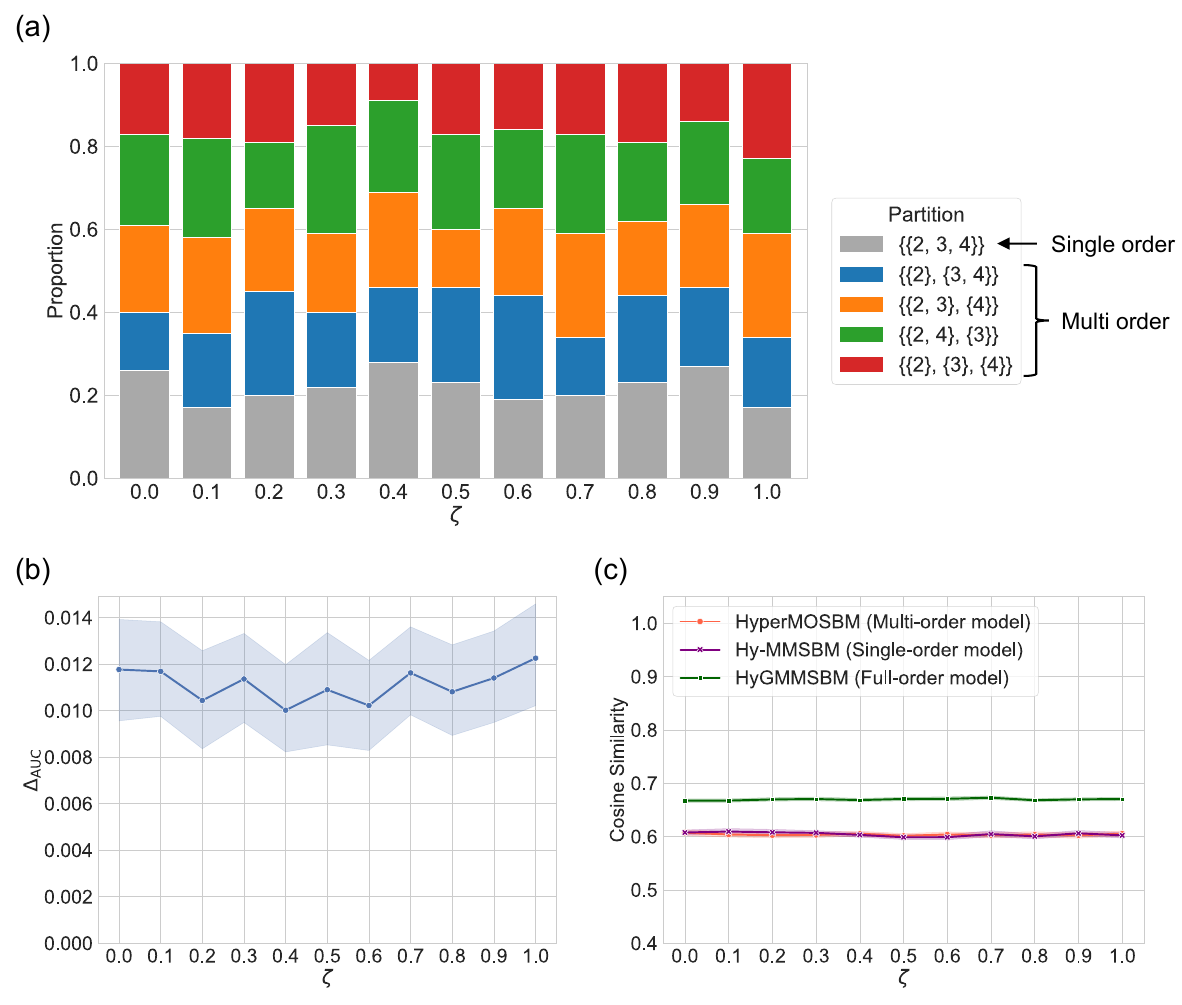}
\caption{
Results on synthetic hypergraphs for $a = 1$.
}
\label{fig:s1}
\end{figure*}

\begin{figure*}[p]
\centering
\includegraphics[width=1.0\textwidth]{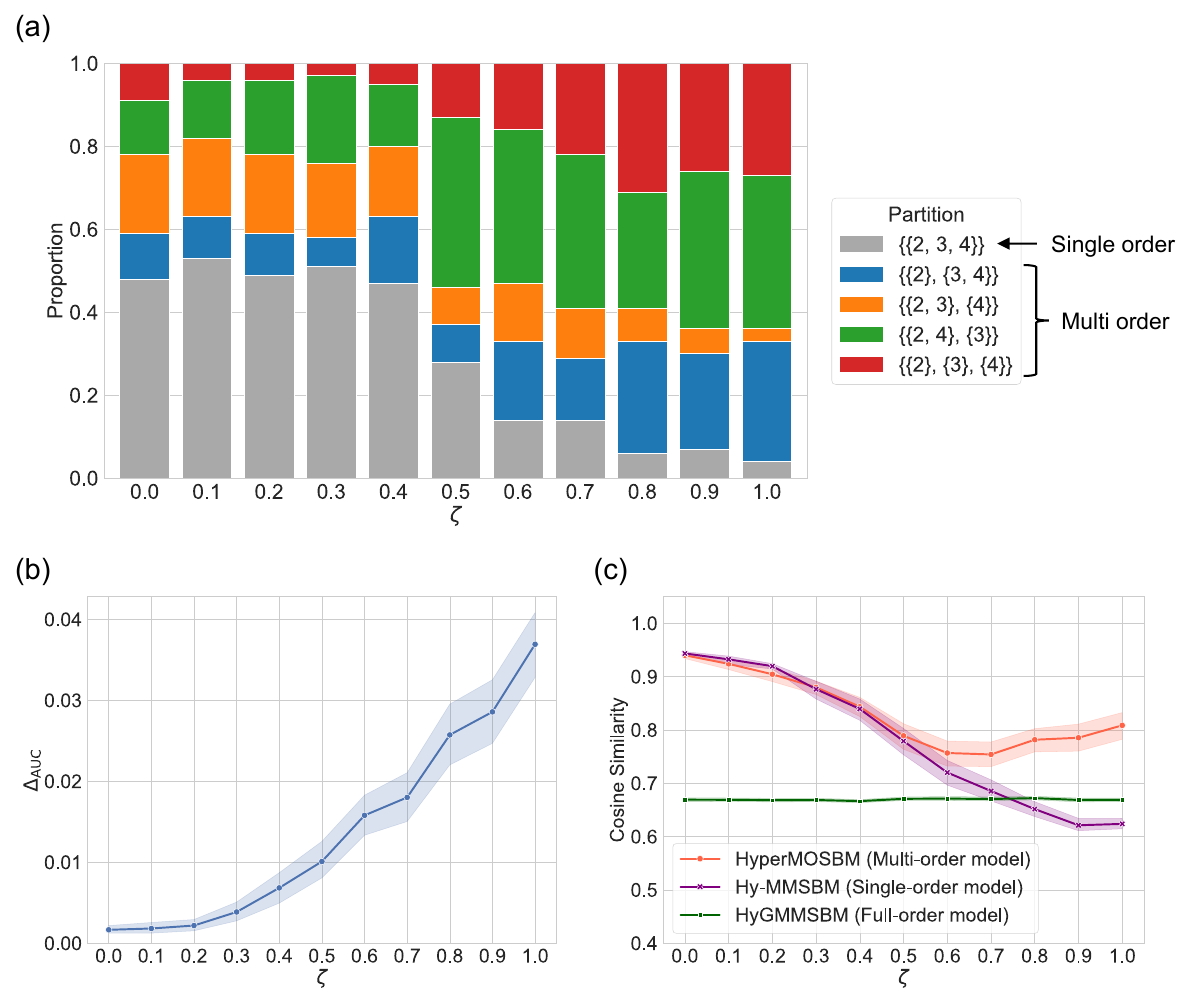}
\caption{
Results on synthetic hypergraphs for $a = 3$.
}
\label{fig:s2}
\end{figure*}

\begin{figure*}[p]
\centering
\includegraphics[width=1.0\textwidth]{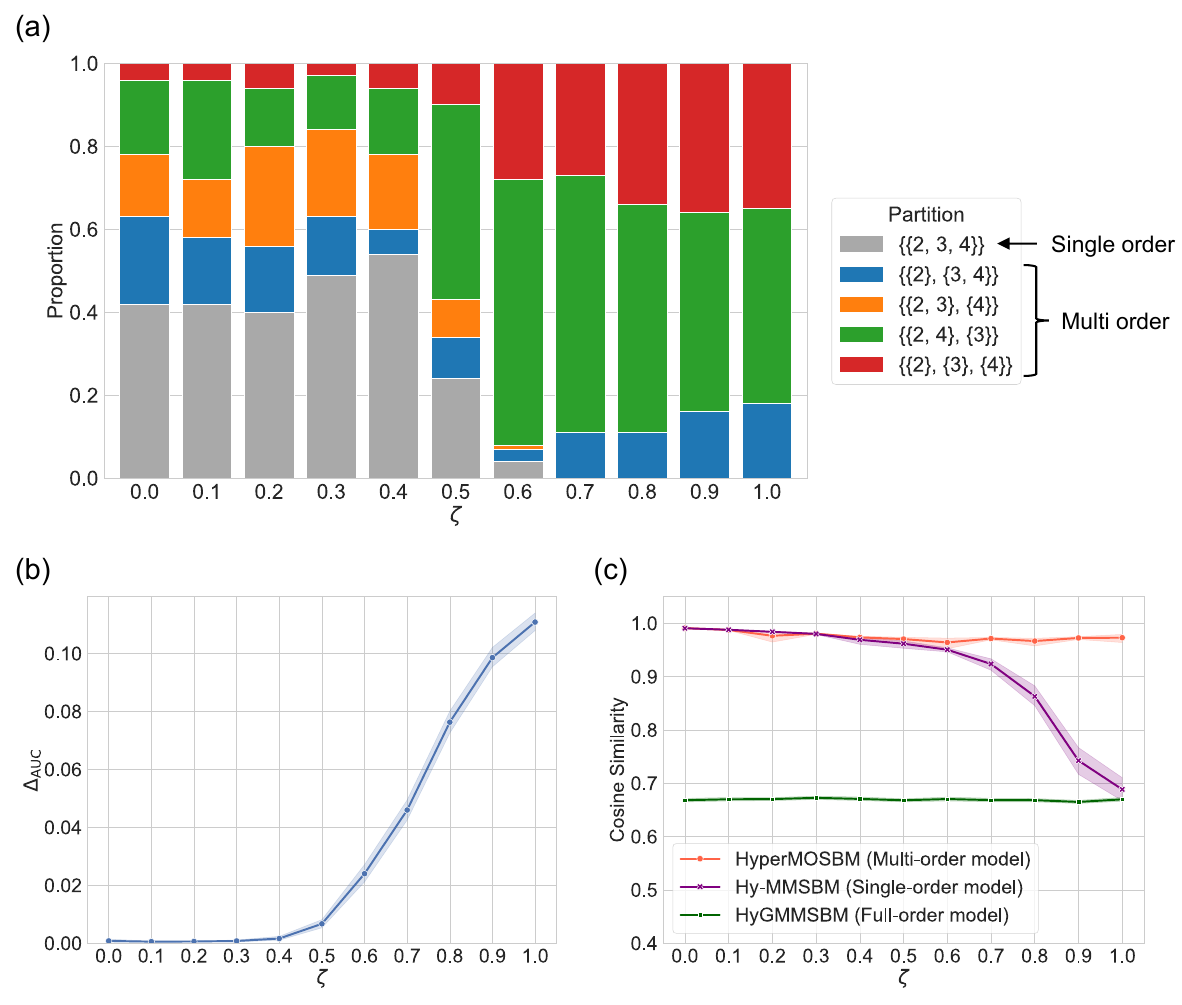}
\caption{
Results on synthetic hypergraphs for $a = 7$.
}
\label{fig:s3}
\end{figure*}

\begin{figure*}[p]
\centering
\includegraphics[width=1.0\textwidth]{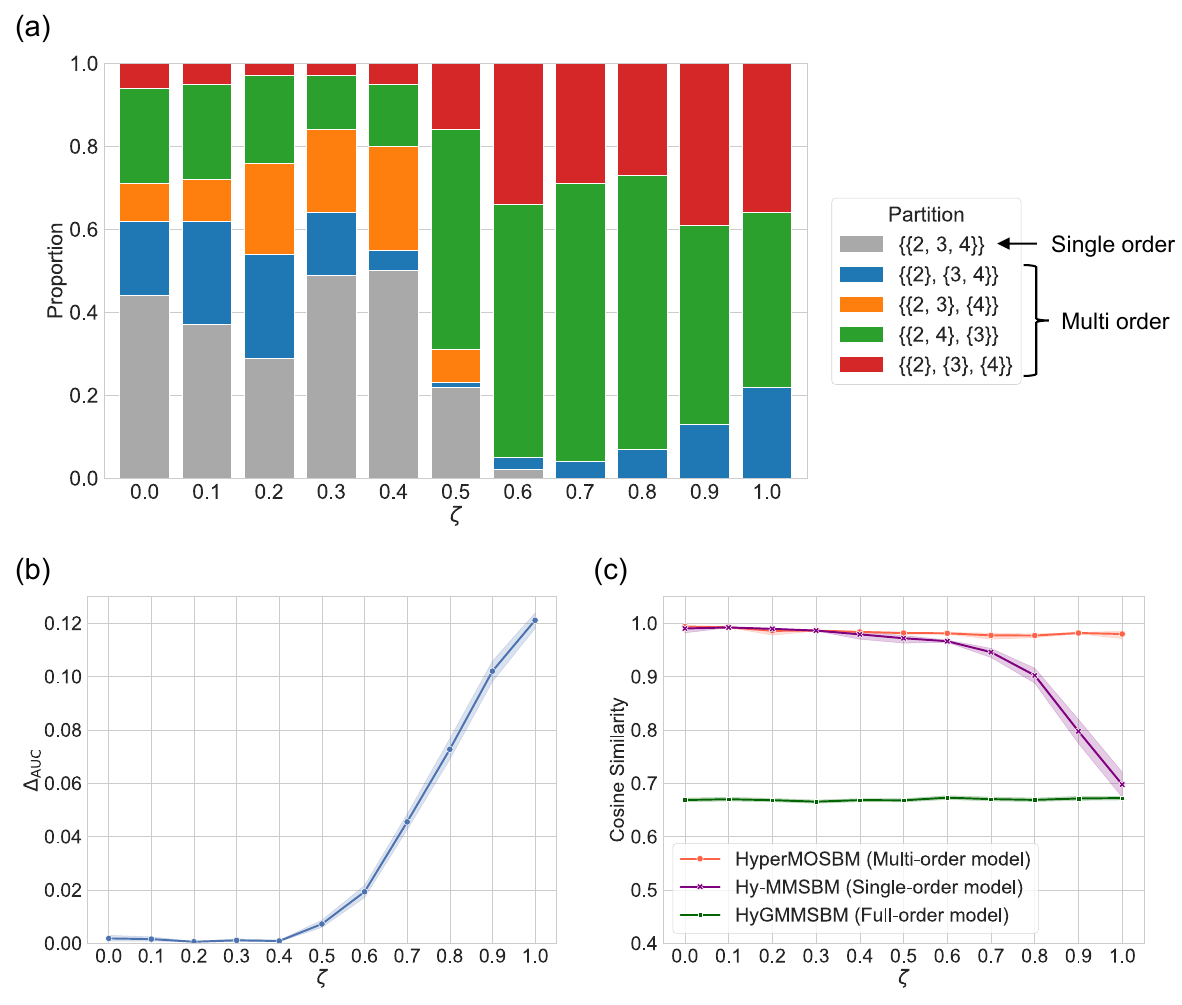}
\caption{
Results on synthetic hypergraphs for $a = 9$.
}
\label{fig:s4}
\end{figure*}

Figures \ref{fig:s1}, \ref{fig:s2}, \ref{fig:s3}, and \ref{fig:s4} show the results in synthetic hypergraphs for $a = 1$, $a = 3$, $a = 7$, and $a = 9$, respectively.

\newpage
\section{Additional results for empirical hypergraphs}

\subsection{Execution times}

Table~\ref{table:s1} provides a detailed comparison of the execution times for the single-order (Hy-MMSBM), multi-order (HyperMOSBM), and full-order (HyGMMSBM) models across all 14 empirical datasets.

\subsection{Final partitions for empirical hypergraphs} 

Table~\ref{table:s2} lists the final partitions of the sets of interaction orders, $\mathcal{P}_{\text{final}}$, for each of the 14 empirical hypergraphs.

\subsection{Inferred affinity matrices for the high-school data}

\begin{figure*}[p]
\centering
\includegraphics[width=1.0\textwidth]{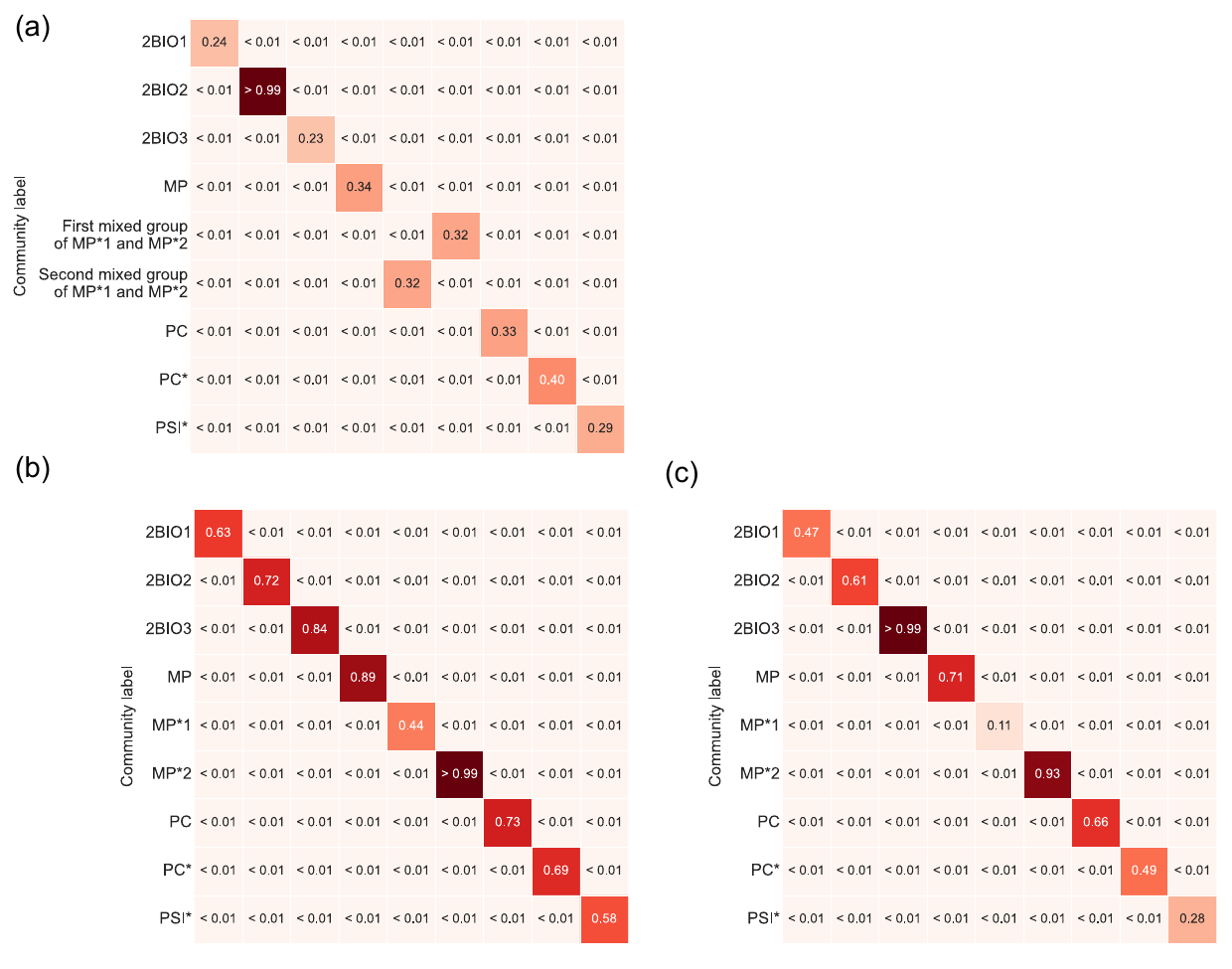}
\caption{
Inferred affinity matrices for the high-school data.
(a) The single affinity matrix inferred by the single-order model. 
(b) The affinity matrix for the hyperedge sizes in $\{2, 4, 5\}$, inferred by our multi-order model.
(c) The affinity matrix for the hyperedge sizes in $\{3\}$, inferred by our multi-order model. 
}
\label{fig:s5}
\end{figure*}

We present the inferred affinity matrices for the high-school data.
Figure~\ref{fig:s5}(a) shows the affinity matrix inferred by the single-order model. 
Figures~\ref{fig:s5}(b) and \ref{fig:s5}(c) show the two distinct affinity matrices inferred by our multi-order model, corresponding to the optimal partition $\mathcal{P}_{\text{final}} = \{\{2, 4, 5\}, \{3\}\}$.
For visualization purposes, all entries in each matrix were normalized by the maximum entry within that matrix.

\subsection{Lists of papers in the inferred NLP and IR communities}

Table~\ref{table:s3} presents examples from the 115 papers classified as representative of the NLP and IR community (i.e., their membership probabilities for that community were at least 0.9) by both models. 
We selected these papers based on their high citation counts and grouped them into five themes.
Tables~\ref{table:s4} and \ref{table:s5} list the papers classified as representative of the NLP and IR community exclusively by the single-order and multi-order models, respectively. 
Note that for papers with identical titles and publication years but different identifiers in the OpenAlex database, we merged them into a single entry and averaged their membership probabilities.

\newpage
\begin{table*}[h]
\caption{Execution time for the single-order model (Hy-MMSBM), multi-order model (HyperMOSBM), and full-order model (HyGMMSBM).}
\label{table:s1}
\begin{center}
\begin{tabular}{lccc}
\toprule
Data & Hy-MMSBM & HyperMOSBM & HyGMMSBM \\ \midrule
high-school & 2.2 min & 11.0 min & 7.2 h \\ \addlinespace
primary-school & 3.7 min & 23.6 min & > 24 h \\ \addlinespace
hospital-lyon & 37.4 s & 6.7 min & 8.3 min \\ \addlinespace
invs13 & 22.0 s & 23.8 s & 2.7 min \\ \addlinespace
invs15 & 1.4 min & 3.6 min & 3.8 h \\ \addlinespace
justice & 58.0 s & 6.4 min & 39.1 min \\ \addlinespace
walmart & 1.3 min & 7.3 min & > 24 h \\ \addlinespace
house-committees & 27.7 s & 2.2 h & > 24 h \\ \addlinespace
senate-committees & 18.1 s & 22.7 min & > 24 h \\ \addlinespace
biochem-cocitations & 28.1 min & 2.0 h & > 24 h \\ \addlinespace
cs-cocitations & 32.7 min & 2.1 h & > 24 h \\ \addlinespace
math-cocitations & 4.8 min & 19.4 min & > 24 h \\ \addlinespace
neuro-cocitations & 5.0 min & 29.3 min & > 24 h \\ \addlinespace
physics-cocitations & 18.1 min & 2.1 h & > 24 h \\ \addlinespace
\bottomrule
\end{tabular}
\end{center}
\end{table*}

\newpage
\begin{table*}[h]
\caption{Final partitions of the sets of interaction orders ($\mathcal{P}_{\text{final}}$) for empirical hypergraphs. }
\label{table:s2}
\begin{center}
\begin{tabular}{ll}
\toprule
Data & $\mathcal{P}_{\text{final}}$ \\ \midrule
high-school & $\{\{2, 4, 5\}, \{3\}\}$ \\ \addlinespace
primary-school & $\{\{2, 3, 5\}, \{4\}\}$ \\ \addlinespace
hospital-lyon & $\{\{2, 4\}, \{3, 5\}\}$ \\ \addlinespace
invs13 & $\{\{2, 3, 4\}\}$ \\ \addlinespace
invs15 & $\{\{2, 4\}, \{3\}\}$ \\ \addlinespace
justice & $\{\{2, \ldots, 9\}\}$ \\ \addlinespace
walmart & $\{\{2, 3, 4\}, \{5, \ldots, 11\}\}$ \\ \addlinespace
house-committees & $\{\{2, \ldots, 16\}, \{17, \ldots, 43\}, \{44, \ldots, 81\}\}$ \\ \addlinespace
senate-committees & $\{\{2, \ldots, 14\}, \{15, \ldots, 31\}\}$ \\ \addlinespace
biochem-cocitations & $\{\{2, 3, 4\}, \{5, \ldots, 17\}\}$ \\ \addlinespace
cs-cocitations & $\{\{2\}, \{3, \ldots, 36\}\}$ \\ \addlinespace
math-cocitations & $\{\{2\}, \{3, \ldots, 13\}\}$ \\ \addlinespace
neuro-cocitations & $\{\{2\}, \{3\}, \{4, \ldots, 50\}\}$ \\ \addlinespace
physics-cocitations & $\{\{2\}, \{3\}, \{4, \ldots, 28\}\}$ \\ \addlinespace
\bottomrule
\end{tabular}
\end{center}
\end{table*}

\newpage
\begin{table*}[h]
\caption{Examples from the 115 papers identified by both models as representative of the NLP and IR community.}
\label{table:s3}
\begin{center}
\begin{tabular}{l}
\toprule
{\bf Statistical NLP} \\
``A statistical approach to machine translation'' (1990). \\ \addlinespace
``The mathematics of statistical machine translation: Parameter estimation'' (1993). \\ \addlinespace
``Accurate unlexicalized parsing'' (2003). \\ \addlinespace
\midrule
{\bf Foundational IR} \\
``A vector space model for automatic indexing'' (1975). \\ \addlinespace
``The probability ranking principle in IR'' (1977). \\ \addlinespace
``A language modeling approach to information retrieval'' (1998). \\ \addlinespace
\midrule
{\bf Representation Learning for NLP} \\
``Distributed representations of words and phrases and their compositionality'' (2013). \\ \addlinespace
``Deep contextualized word representations'' (2018). \\ \addlinespace
``Transformers: State-of-the-art natural language processing'' (2020). \\ \addlinespace
\midrule
{\bf Text Mining and Analytics} \\
``Mining and summarizing customer reviews'' (2004). \\ \addlinespace
``TextRank: Bringing order into text'' (2004). \\ \addlinespace
``Sentiment analysis algorithms and applications: A survey'' (2014). \\ \addlinespace
\midrule
{\bf Resources and Evaluation} \\
``BLEU: A method for automatic evaluation of machine translation'' (2002). \\ \addlinespace
``NLTK: The Natural Language Toolkit'' (2002). \\ \addlinespace
``The Stanford CoreNLP Natural Language Processing Toolkit'' (2014). \\ \addlinespace
\bottomrule
\end{tabular}
\end{center}
\end{table*}

\newpage
\begin{longtable}{p{14cm} r r}
\caption{List of the 44 papers classified as representative of the NLP and IR community exclusively by the single-order model (Hy-MMSBM). We denote by $p_{\text{single}}$ and $p_{\text{multi}}$ the membership probabilities in the single-order and multi-order models, respectively.} \label{table:s4} \\
\toprule
{\bf Paper} & $p_{\text{single}}$ & $p_{\text{multi}}$ \\
\midrule
\endfirsthead
\hline
{\bf Paper} & $p_{\text{single}}$ & $p_{\text{multi}}$ \\
\hline
\endhead
\multicolumn{3}{l}{\bf Foundational IR} \\
\addlinespace
``A statistical interpretation of term specificity and its application in retrieval'' (1972) & 0.96 & 0.89 \\
\addlinespace
``Term-weighting approaches in automatic text retrieval'' (1988) & 0.93 & 0.84 \\
\addlinespace
``Reexamining the cluster hypothesis: Scatter/gather on retrieval results'' (1996) & 0.94 & 0.86 \\
\addlinespace
``Syntactic clustering of the web'' (1997) & 0.97 & 0.72 \\
\addlinespace
``The anatomy of a large-scale hypertextual web search engine'' (1998) & 0.97 & 0.90 \\
\addlinespace
``The PageRank citation ranking: Bringing order to the web'' (1999) & 0.95 & 0.78 \\
\addlinespace
``Modern information retrieval: A brief overview'' (2001) & 0.97 & 0.87 \\
\addlinespace
``Optimizing search engines using clickthrough data'' (2002) & 0.95 & 0.89 \\
\addlinespace
\midrule
\multicolumn{3}{l}{\bf Statistical NLP} \\
\addlinespace
``Indexing by latent semantic analysis'' (1990) & 0.92 & 0.86 \\
\addlinespace
``Unsupervised word sense disambiguation rivaling supervised methods'' (1995) & 0.97 & 0.86 \\
\addlinespace
``A maximum entropy approach to natural language processing'' (1996) & 0.93 & 0.81 \\
\addlinespace
``A comparison of event models for naive bayes text classification'' (1998) & 0.98 & 0.74 \\
\addlinespace
``An empirical study of smoothing techniques for language modeling'' (1999) & 0.94 & 0.82 \\
\addlinespace
``Latent dirichlet allocation'' (2003) & 0.90 & 0.84 \\
\addlinespace
``Head-driven statistical models for natural language parsing'' (2003) & 0.93 & 0.87 \\
\addlinespace
``RCV1: A New Benchmark Collection for Text Categorization Research'' (2004) & 0.92 & 0.70 \\
\midrule
\multicolumn{3}{l}{\bf Distributed Systems and Big Data} \\
\addlinespace
``The Google file system'' (2003) & 0.97 & $<$ 0.01 \\
\addlinespace
``Ceph: A scalable, high-performance distributed file system'' (2006) & $>$ 0.99 & $<$ 0.01 \\
\addlinespace
``Bigtable: A distributed storage system for structured data'' (2006) & 0.94 & $<$ 0.01 \\
\addlinespace
``Dryad: distributed data-parallel programs from sequential building blocks'' (2007) & $>$ 0.99 & $<$ 0.01 \\
\addlinespace
``Improving MapReduce performance in heterogeneous environments'' (2008) & $>$ 0.99 & $<$ 0.01 \\
\addlinespace
``Bigtable: A distributed storage system for structured data'' (2008) & 0.98 & $<$ 0.01 \\
\addlinespace
``Pig latin: a not-so-foreign language for data processing'' (2008) & $>$ 0.99 & $<$ 0.01 \\
\addlinespace
``Quincy: fair scheduling for distributed computing clusters'' (2009) & $>$ 0.99 & $<$ 0.01 \\
\addlinespace
``Hive: a warehousing solution over a map-reduce framework'' (2009) & $>$ 0.99 & $<$ 0.01 \\
\addlinespace
``HadoopDB: an architectural hybrid of MapReduce and DBMS technologies for analytical workloads'' (2009) & $>$ 0.99 & $<$ 0.01 \\
\addlinespace
``MapReduce: a flexible data processing tool'' (2009) & $>$ 0.99 & $<$ 0.01 \\
\addlinespace
``Twister: a runtime for iterative mapreduce'' (2010) & $>$ 0.99 & $<$ 0.01 \\
\addlinespace
``Delay scheduling: a simple technique for achieving locality and fairness in cluster scheduling'' (2010) & 0.93 & $<$ 0.01 \\
\addlinespace
``The hadoop distributed file system'' (2010) & 0.95 & $<$ 0.01 \\
\addlinespace
``Spark: Cluster computing with working sets'' (2010) & $>$ 0.99 & 0.03 \\
\addlinespace
``Mesos: A Platform for Fine-Grained Resource Sharing in the Data Center'' (2011) & $>$ 0.99 & $<$ 0.01 \\
\addlinespace
``Resilient Distributed Datasets: A Fault-Tolerant Abstraction for In-Memory Cluster Computing'' (2012) & $>$ 0.99 & $<$ 0.01 \\
\addlinespace
``Apache Hadoop YARN: yet another resource negotiator'' (2013) & $>$ 0.99 & $<$ 0.01 \\
\addlinespace
``Large-scale cluster management at Google with Borg'' (2015) & $>$ 0.99 & $<$ 0.01 \\
\addlinespace
``Apache Spark: a unified engine for big data processing'' (2016) & 0.92 & $<$ 0.01 \\
\addlinespace
\midrule
\multicolumn{3}{l}{\bf Complex Systems} \\
\addlinespace
``Concentration wave propagation in two-dimensional liquid-phase self-oscillating system'' (1970) & $>$ 0.99 & $<$ 0.01 \\
\addlinespace
``Oscillations in chemical systems. II. Thorough analysis of temporal oscillation in the bromate-cerium-malonic acid system'' (1972) & $>$ 0.99 & $<$ 0.01 \\
\addlinespace
``A set of measures of centrality based on betweenness'' (1977) & $>$ 0.99 & 0.48 \\
\addlinespace
``Master stability functions for synchronized coupled systems'' (1998) & 0.90 & $<$ 0.01 \\
\addlinespace
``Synchronization in small-world systems'' (2002) & $>$ 0.99 & $<$ 0.01 \\
\addlinespace
``The synchronization of chaotic systems'' (2002) & 0.93 & $<$ 0.01 \\
\addlinespace
``Synchronization in scale-free dynamical networks: robustness and fragility'' (2002) & $>$ 0.99 & $<$ 0.01 \\
\addlinespace
``Chimera states for coupled oscillators'' (2004) & $>$ 0.99 & $<$ 0.01 \\
\addlinespace
\bottomrule
\end{longtable}

\newpage
\begin{longtable}{p{14cm} r r}
\caption{List of the 29 papers classified as representative of the NLP and IR community exclusively by the multi-order model (HyperMOSBM). We denote by $p_{\text{single}}$ and $p_{\text{multi}}$ the membership probabilities in the single-order and multi-order models, respectively.} \label{table:s5} \\
\toprule
{\bf Paper} & $p_{\text{single}}$ & $p_{\text{multi}}$ \\
\midrule
\endfirsthead
\hline
{\bf Paper} & $p_{\text{single}}$ & $p_{\text{multi}}$ \\
\hline
\endhead
\multicolumn{3}{l}{\bf Neural Networks and Representation Learning} \\
\addlinespace
``An information-theoretic definition of similarity'' (1998) & 0.88 & $>$ 0.99 \\
\addlinespace
``A neural probabilistic language model'' (2003) & 0.85 & 0.92 \\
\addlinespace
``Glove: Global vectors for word representation'' (2014) & 0.82 & 0.91 \\
\addlinespace
``Effective approaches to attention-based neural machine translation'' (2015) & 0.80 & 0.94 \\
\addlinespace
``Neural machine translation by jointly learning to align and translate'' (2015) & 0.83 & 0.93 \\
\addlinespace
``Hierarchical attention networks for document classification'' (2016) & 0.83 & 0.95 \\
\addlinespace
``Wide \& Deep Learning for Recommender Systems'' (2016) & $<$ 0.01 & 0.95 \\
\addlinespace
``Heterogeneous information network embedding for recommendation'' (2018) & 0.36 & $>$ 0.99 \\
\addlinespace
\midrule
\multicolumn{3}{l}{\bf Semantics, Knowledge, and Web Mining} \\
\addlinespace
``Introduction to WordNet: An on-line lexical database'' (1990) & 0.86 & 0.98 \\
\addlinespace
``Automatic acquisition of hyponyms from large text corpora'' (1992) & 0.89 & 0.92 \\
\addlinespace
``Verb semantics and lexical selection'' (1994) & 0.88 & $>$ 0.99 \\
\addlinespace
``WordNet: a lexical database for English'' (1995) & 0.88 & 0.93 \\
\addlinespace
``Web mining research: A survey'' (2000) & 0.85 & 0.94 \\
\addlinespace
``WordNet: An electronic lexical database'' (2000) & 0.88 & 0.91 \\
\addlinespace
``Placing search in context: The concept revisited'' (2002) & 0.88 & $>$ 0.99 \\
\addlinespace
``Exploratory search: from finding to understanding'' (2006) & 0.78 & 0.94 \\
\addlinespace
``Usage patterns of collaborative tagging systems'' (2006) & 0.89 & $>$ 0.99 \\
\addlinespace
``Collaborative topic modeling for recommending scientific articles'' (2011) & 0.67 & $>$ 0.99 \\
\addlinespace
``Collaborative knowledge base embedding for recommender systems'' (2016) & 0.47 & $>$ 0.99 \\
\addlinespace
\midrule
\multicolumn{3}{l}{\bf Computer Security} \\
\addlinespace
``Computer security threat monitoring and surveillance'' (1980) & $<$ 0.01 & $>$ 0.99 \\
\addlinespace
``A sense of self for unix processes'' (1996) & $<$ 0.01 & $>$ 0.99 \\
\addlinespace
``Intrusion detection using sequences of system calls'' (1998) & $<$ 0.01 & $>$ 0.99 \\
\addlinespace
``Testing Intrusion detection systems: a critique of the 1998 and 1999 DARPA intrusion detection system evaluations as performed by Lincoln Laboratory'' (2000) & $<$ 0.01 & $>$ 0.99 \\
\addlinespace
``Outside the closed world: On using machine learning for network intrusion detection'' (2010) & $<$ 0.01 & $>$ 0.99 \\
\addlinespace
``Toward developing a systematic approach to generate benchmark datasets for intrusion detection'' (2012) & $<$ 0.01 & $>$ 0.99 \\
\addlinespace
``UNSW-NB15: a comprehensive data set for network intrusion detection systems (UNSW-NB15 network data set)'' (2015) & $<$ 0.01 & 0.93 \\
\addlinespace
``A deep learning approach for intrusion detection using recurrent neural networks'' (2017) & $<$ 0.01 & 0.93 \\
\addlinespace
``A deep learning approach to network intrusion detection'' (2018) & $<$ 0.01 & $>$ 0.99 \\
\addlinespace
``Deep learning approach for intelligent intrusion detection system'' (2019) & $<$ 0.01 & $>$ 0.99 \\
\addlinespace
\bottomrule
\end{longtable}

\newpage
\section{Implementation details}

In our implementation, the EM algorithm iterates $N_{\text{I}}$ times for a given initial configuration of the latent parameters.
Furthermore, to mitigate the convergence to a poor local maximum of the likelihood, we perform $N_{\text{R}}$ independent trials, each starting from a different randomly initialized configuration of the parameters \cite{newman2016, debacco2017, contisciani2022, nakajima2025}. 
We initialize all $N_{\text{R}}$ trials uniformly at random. 
Among the $N_{\text{R}}$ inferred parameter sets, we select the one that yields the highest final log-likelihood value. 
To generate the benchmark results (Table 2), we set $N_{\text{I}} = 500$ and $N_{\text{R}} = 10$.
In the further analyses of the `high-school' and `cs-cocitations' networks,  we set $N_{\text{I}} = 500$ and $N_{\text{R}} = 10^3$.
The pseudocode for HyperMOSBM is provided in Algorithm \ref{alg:s1}.

We used the implementation of the single-order model provided by the `hypergraphx' library (version 1.7.8) \cite{lotito2023}.
We used the implementation of the full-order model obtained from the authors' original repository at \url{https://github.com/seeslab/HyGMMSBM} (accessed October 2025).
We configured both models to run for the same number of iterations and random initializations as our multi-order model.
All three models are implemented in Python.

\begin{algorithm}[h]
\caption{HyperMOSBM}
\label{alg:s1}
\begin{algorithmic}[1]
\REQUIRE Hypergraph: $\mathcal{A}$; number of communities: $K$; partition of the set of hyperedges' sizes: $\mathcal{P}$.
\ENSURE Inferred variables: $\hat{\bm{\theta}}$.
\STATE{$L_{\text{best}} = - \infty$}
\STATE{$\hat{\bm{\theta}} =$ None}
\FOR{$r=1, \ldots, N_{\text{R}}$}
\STATE{Initialize $\hat{\bm{U}}$ and $\{\hat{\bm{W}}^{(l)}\}_{l=1}^L$.}
\FOR{$i=1, \ldots, N_{\text{I}}$}
\STATE{Calculate $\rho$ using Eq.~(13).}
\STATE{Update $\hat{\bm{U}}$ using Eq.~(16).}
\FOR{$l=1, \ldots, L$}
\STATE{Update $\hat{\bm{W}}^{(l)}$ using Eq.~(18).}
\ENDFOR
\ENDFOR
\STATE{$L = \mathcal{L}(\hat{\bm{U}}, \{\hat{\bm{W}}^{(l)}\}_{l=1}^L\ |\ \mathcal{A})$ (Eq.~(4))}
\IF{$L > L_{\text{best}}$}
\STATE{$L_{\text{best}} \leftarrow L$}
\STATE{$\hat{\bm{\theta}} \leftarrow (\hat{\bm{U}}, \{\hat{\bm{W}}^{(l)}\}_{l=1}^L)$}
\ENDIF
\ENDFOR
\RETURN{$\hat{\bm{\theta}}$}
\end{algorithmic}
\end{algorithm}

\renewcommand{\refname}{Supplementary References}


\begin{thebibliography}{100}

\bibitem{boccaletti2006}
S.~Boccaletti, V.~Latora, Y.~Moreno, M.~Chavez, and D.-U. Hwang.
\newblock Complex networks: Structure and dynamics.
\newblock {\em Physics Reports}, 424:175--308, 2006.

\bibitem{newman2018}
M.~E.~J. Newman.
\newblock {\em Networks. 2nd ed.}
\newblock Oxford university press, 2018.

\bibitem{milo2002}
R.~Milo, S.~Shen-Orr, S.~Itzkovitz, N.~Kashtan, D.~Chklovskii, and U.~Alon.
\newblock Network motifs: Simple building blocks of complex networks.
\newblock {\em Science}, 298:824--827, 2002.

\bibitem{girvan2002}
M.~Girvan and M.~E.~J. Newman.
\newblock Community structure in social and biological networks.
\newblock {\em Proceedings of the National Academy of Sciences}, 99:7821--7826, 2002.

\bibitem{colizza2006}
V.~Colizza, A.~Flammini, M.~A. Serrano, and A.~Vespignani.
\newblock Detecting rich-club ordering in complex networks.
\newblock {\em Nature Physics}, 2:110--115, 2006.

\bibitem{clauset2008}
Aaron Clauset, Cristopher Moore, and M.~E.~J. Newman.
\newblock Hierarchical structure and the prediction of missing links in networks.
\newblock {\em Nature}, 453:98--101, 2008.

\bibitem{lotito2022}
Quintino~Francesco Lotito, Federico Musciotto, Alberto Montresor, and Federico Battiston.
\newblock Higher-order motif analysis in hypergraphs.
\newblock {\em Communications Physics}, 5:79, 2022.

\bibitem{nakajima2023}
Kazuki Nakajima, Kazuyuki Shudo, and Naoki Masuda.
\newblock Higher-order rich-club phenomenon in collaborative research grant networks.
\newblock {\em Scientometrics}, 128:2429--2446, 2023.

\bibitem{fortunato2016}
Santo Fortunato and Darko Hric.
\newblock Community detection in networks: A user guide.
\newblock {\em Physics Reports}, 659:1--44, 2016.

\bibitem{ravasz2002}
E.~Ravasz, A.~L. Somera, D.~A. Mongru, Z.~N. Oltvai, and A.-L. Barab^^c3^^a1si.
\newblock Hierarchical organization of modularity in metabolic networks.
\newblock {\em Science}, 297:1551--1555, 2002.

\bibitem{guimera2005}
Roger Guimer{\`a} and Lu{\'\i}s~A. Nunes~Amaral.
\newblock Functional cartography of complex metabolic networks.
\newblock {\em Nature}, 433:895--900, 2005.

\bibitem{zachary1977}
Wayne~W. Zachary.
\newblock An information flow model for conflict and fission in small groups.
\newblock {\em Journal of Anthropological Research}, 33:452--473, 1977.

\bibitem{lusseau2003}
David Lusseau, Karsten Schneider, Oliver~J. Boisseau, Patti Haase, Elisabeth Slooten, and Steve~M. Dawson.
\newblock The bottlenose dolphin community of doubtful sound features a large proportion of long-lasting associations.
\newblock {\em Behavioral Ecology and Sociobiology}, 54:396--405, 2003.

\bibitem{pastorsatorras2015}
Romualdo Pastor-Satorras, Claudio Castellano, Piet Van~Mieghem, and Alessandro Vespignani.
\newblock Epidemic processes in complex networks.
\newblock {\em Reviews of Modern Physics}, 87:925--979, 2015.

\bibitem{stehle2011}
Juliette Stehl^^c3^^a9, Nicolas Voirin, Alain Barrat, Ciro Cattuto, Lorenzo Isella, Jean-Fran^^c3^^a7ois Pinton, Marco Quaggiotto, Wouter Van~den Broeck, Corinne R^^c3^^a9gis, Bruno Lina, and Philippe Vanhems.
\newblock High-resolution measurements of face-to-face contact patterns in a primary school.
\newblock {\em PLOS ONE}, 6:e23176, 2011.

\bibitem{mastrandrea2015}
Rossana Mastrandrea, Julie Fournet, and Alain Barrat.
\newblock Contact patterns in a high school: A comparison between data collected using wearable sensors, contact diaries and friendship surveys.
\newblock {\em PLOS ONE}, 10:e0136497, 2015.

\bibitem{newman2001}
M.~E.~J. Newman.
\newblock The structure of scientific collaboration networks.
\newblock {\em Proceedings of the National Academy of Sciences}, 98:404--409, 2001.

\bibitem{patania2017}
Alice Patania, Giovanni Petri, and Francesco Vaccarino.
\newblock The shape of collaborations.
\newblock {\em EPJ Data Science}, 6(1):18, 2017.

\bibitem{wong2008}
Philip Wong, Sonja Althammer, Andrea Hildebrand, Andreas Kirschner, Philipp Pagel, Bernd Geissler, Pawel Smialowski, Florian Bl{\"o}chl, Matthias Oesterheld, Thorsten Schmidt, Normann Strack, Fabian~J. Theis, Andreas Ruepp, and Dmitrij Frishman.
\newblock An evolutionary and structural characterization of mammalian protein complex organization.
\newblock {\em BMC Genomics}, 9:629, 2008.

\bibitem{gaudelet2018}
Thomas Gaudelet, No^^c3^^abl Malod-Dognin, and Nata^^c5^^a1a Pr^^c5^^beulj.
\newblock Higher-order molecular organization as a source of biological function.
\newblock {\em Bioinformatics}, 34:i944--i953, 2018.

\bibitem{giusti2016}
Chad Giusti, Robert Ghrist, and Danielle~S. Bassett.
\newblock Two's company, three (or more) is a simplex.
\newblock {\em Journal of Computational Neuroscience}, 41:1--14, 2016.

\bibitem{varley2023}
Thomas~F. Varley, Maria Pope, null, null, and Olaf Sporns.
\newblock Partial entropy decomposition reveals higher-order information structures in human brain activity.
\newblock {\em Proceedings of the National Academy of Sciences}, 120:e2300888120, 2023.

\bibitem{battiston2020}
Federico Battiston, Giulia Cencetti, Iacopo Iacopini, Vito Latora, Maxime Lucas, Alice Patania, Jean-Gabriel Young, and Giovanni Petri.
\newblock Networks beyond pairwise interactions: Structure and dynamics.
\newblock {\em Physics Reports}, 874:1--92, 2020.

\bibitem{battiston2021}
Federico Battiston, Enrico Amico, Alain Barrat, Ginestra Bianconi, Guilherme Ferraz~de Arruda, Benedetta Franceschiello, Iacopo Iacopini, Sonia K{\'e}fi, Vito Latora, Yamir Moreno, Micah~M. Murray, Tiago~P. Peixoto, Francesco Vaccarino, and Giovanni Petri.
\newblock The physics of higher-order interactions in complex systems.
\newblock {\em Nature Physics}, 17:1093--1098, 2021.

\bibitem{boccaletti2023}
S.~Boccaletti, P.~{De Lellis}, C.I. {del Genio}, K.~Alfaro-Bittner, R.~Criado, S.~Jalan, and M.~Romance.
\newblock The structure and dynamics of networks with higher order interactions.
\newblock {\em Physics Reports}, 1018:1--64, 2023.
\newblock The structure and dynamics of networks with higher order interactions.

\bibitem{antelmi2023}
Alessia Antelmi, Gennaro Cordasco, Mirko Polato, Vittorio Scarano, Carmine Spagnuolo, and Dingqi Yang.
\newblock A survey on hypergraph representation learning.
\newblock {\em ACM Computing Surveys}, 56, 2023.

\bibitem{majhi2022}
Soumen Majhi, Matja^^c5^^be Perc, and Dibakar Ghosh.
\newblock Dynamics on higher-order networks: A review.
\newblock {\em Journal of The Royal Society Interface}, 19:20220043, 2022.

\bibitem{lee2025}
Geon Lee, Fanchen Bu, Tina Eliassi-Rad, and Kijung Shin.
\newblock A survey on hypergraph mining: Patterns, tools, and generators.
\newblock {\em ACM Computing Surveys}, 57, 2025.

\bibitem{lotito2023}
Quintino~Francesco Lotito, Martina Contisciani, Caterina De~Bacco, Leonardo Di~Gaetano, Luca Gallo, Alberto Montresor, Federico Musciotto, Nicol^^c3^^b2 Ruggeri, and Federico Battiston.
\newblock Hypergraphx: {A} library for higher-order network analysis.
\newblock {\em Journal of Complex Networks}, 11:cnad019, 2023.

\bibitem{landry2023}
Nicholas~W. Landry, Maxime Lucas, Iacopo Iacopini, Giovanni Petri, Alice Schwarze, Alice Patania, and Leo Torres.
\newblock {XGI: A Python package for higher-order interaction networks}.
\newblock {\em Journal of Open Source Software}, 8:5162, 2023.

\bibitem{airoldi2008}
Edo~M Airoldi, David Blei, Stephen Fienberg, and Eric Xing.
\newblock Mixed membership stochastic blockmodels.
\newblock In {\em Advances in Neural Information Processing Systems}, volume~21, 2008.

\bibitem{karrer2011}
Brian Karrer and M.~E.~J. Newman.
\newblock Stochastic blockmodels and community structure in networks.
\newblock {\em Physical Review E}, 83:016107, 2011.

\bibitem{lee2019}
Clement Lee and Darren~J. Wilkinson.
\newblock A review of stochastic block models and extensions for graph clustering.
\newblock {\em Applied Network Science}, 4:122, 2019.

\bibitem{chodrow2021}
Philip~S. Chodrow, Nate Veldt, and Austin~R. Benson.
\newblock Generative hypergraph clustering: From blockmodels to modularity.
\newblock {\em Science Advances}, 7:eabh1303, 2021.

\bibitem{contisciani2022}
Martina Contisciani, Federico Battiston, and Caterina De~Bacco.
\newblock Inference of hyperedges and overlapping communities in hypergraphs.
\newblock {\em Nature Communications}, 13:7229, 2022.

\bibitem{sales2023}
Marta Sales-Pardo, Aleix Marin^^c3^^a9-Tena, and Roger Guimer^^c3^^a0.
\newblock Hyperedge prediction and the statistical mechanisms of higher-order and lower-order interactions in complex networks.
\newblock {\em Proceedings of the National Academy of Sciences}, 120:e2303887120, 2023.

\bibitem{brusa2024}
Luca Brusa and Catherine Matias.
\newblock Model-based clustering in simple hypergraphs through a stochastic blockmodel.
\newblock {\em Scandinavian Journal of Statistics}, 51:1661--1684, 2024.

\bibitem{ruggeri2023}
Nicol^^c3^^b2 Ruggeri, Martina Contisciani, Federico Battiston, and Caterina~De Bacco.
\newblock Community detection in large hypergraphs.
\newblock {\em Science Advances}, 9:eadg9159, 2023.

\bibitem{hood2025}
John Hood, Caterina De~Bacco, and Aaron Schein.
\newblock Broad spectrum structure discovery in large-scale higher-order networks.
\newblock {\em arXiv preprint arXiv:2505.21748}, 2025.

\bibitem{guimera2009}
Roger Guimer^^c3^^a0 and Marta Sales-Pardo.
\newblock Missing and spurious interactions and the reconstruction of complex networks.
\newblock {\em Proceedings of the National Academy of Sciences}, 106:22073--22078, 2009.

\bibitem{aicher2014}
Christopher Aicher, Abigail~Z. Jacobs, and Aaron Clauset.
\newblock Learning latent block structure in weighted networks.
\newblock {\em Journal of Complex Networks}, 3:221--248, 2014.

\bibitem{vallescatala2016}
Toni Vall\`es-Catal\`a, Francesco~A. Massucci, Roger Guimer\`a, and Marta Sales-Pardo.
\newblock Multilayer stochastic block models reveal the multilayer structure of complex networks.
\newblock {\em Physical Review X}, 6:011036, 2016.

\bibitem{ghasemian2020}
Amir Ghasemian, Homa Hosseinmardi, and Aaron Clauset.
\newblock Evaluating overfit and underfit in models of network community structure.
\newblock {\em IEEE Transactions on Knowledge and Data Engineering}, 32:1722--1735, 2020.

\bibitem{ghoshdastidar2017}
Debarghya Ghoshdastidar and Ambedkar Dukkipati.
\newblock Consistency of spectral hypergraph partitioning under planted partition model.
\newblock {\em The Annals of Statistics}, pages 289--315, 2017.

\bibitem{dumitriu2025}
Ioana Dumitriu, Hai-Xiao Wang, and Yizhe Zhu.
\newblock Partial recovery and weak consistency in the non-uniform hypergraph stochastic block model.
\newblock {\em Combinatorics, Probability and Computing}, 34:1^^e2^^80^^9351, 2025.

\bibitem{debacco2017}
Caterina De~Bacco, Eleanor~A. Power, Daniel~B. Larremore, and Cristopher Moore.
\newblock Community detection, link prediction, and layer interdependence in multilayer networks.
\newblock {\em Physical Review E}, 95:042317, 2017.

\bibitem{benson}
Austin~R. Benson.
\newblock \url{https://www.cs.cornell.edu/~arb/data/}, 2025.
\newblock Accessed September 2025.

\bibitem{gemmetto2014}
Valerio Gemmetto, Alain Barrat, and Ciro Cattuto.
\newblock Mitigation of infectious disease at school: targeted class closure vs school closure.
\newblock {\em BMC Infectious Diseases}, 14:695, 2014.

\bibitem{vanhems2013}
Philippe Vanhems, Alain Barrat, Ciro Cattuto, Jean-Fran^^c3^^a7ois Pinton, Nagham Khanafer, Corinne R^^c3^^a9gis, Byeul-a Kim, Brigitte Comte, and Nicolas Voirin.
\newblock Estimating potential infection transmission routes in hospital wards using wearable proximity sensors.
\newblock {\em PLOS ONE}, 8:e73970, 2013.

\bibitem{genois2015}
Mathieu G{\'e}nois, Christian~L Vestergaard, Julie Fournet, Andr{\'e} Panisson, Isabelle Bonmarin, and Alain Barrat.
\newblock Data on face-to-face contacts in an office building suggest a low-cost vaccination strategy based on community linkers.
\newblock {\em Network Science}, 3:326^^e2^^80^^93347, 2015.

\bibitem{genois2018}
Mathieu G{\'e}nois and Alain Barrat.
\newblock Can co-location be used as a proxy for face-to-face contacts?
\newblock {\em EPJ Data Science}, 7:11, 2018.

\bibitem{amburg2020}
Ilya Amburg, Nate Veldt, and Austin Benson.
\newblock Clustering in graphs and hypergraphs with categorical edge labels.
\newblock In {\em Proceedings of The Web Conference 2020}, pages 706--717, 2020.

\bibitem{house_committees}
{Charles Stewart III and Jonathan Woon}.
\newblock {Congressional Committee Assignments, 103rd to 114th Congresses, 1993--2017: House}, 2017.

\bibitem{senate_committees}
{Charles Stewart III and Jonathan Woon}.
\newblock {Congressional Committee Assignments, 103rd to 114th Congresses, 1993--2017: Senate}, 2017.

\bibitem{priem2022}
Jason Priem, Heather Piwowar, and Richard Orr.
\newblock {OpenAlex}: {A} fully-open index of scholarly works, authors, venues, institutions, and concepts.
\newblock {\em arXiv preprint arXiv:2205.01833}, 2022.

\bibitem{trujillo2018}
Caleb~M. Trujillo and Tammy~M. Long.
\newblock Document co-citation analysis to enhance transdisciplinary research.
\newblock {\em Science Advances}, 4:e1701130, 2018.

\bibitem{chakraborty2018}
Tanmoy Chakraborty.
\newblock Role of interdisciplinarity in computer sciences: quantification, impact and life trajectory.
\newblock {\em Scientometrics}, 114:1011--1029, 2018.

\bibitem{peel2017}
Leto Peel, Daniel~B. Larremore, and Aaron Clauset.
\newblock The ground truth about metadata and community detection in networks.
\newblock {\em Science Advances}, 3:e1602548, 2017.

\bibitem{brown2020}
Tom Brown, Benjamin Mann, Nick Ryder, Melanie Subbiah, Jared~D Kaplan, Prafulla Dhariwal, Arvind Neelakantan, Pranav Shyam, Girish Sastry, Amanda Askell, Sandhini Agarwal, Ariel Herbert-Voss, Gretchen Krueger, Tom Henighan, Rewon Child, Aditya Ramesh, Daniel Ziegler, Jeffrey Wu, Clemens Winter, Chris Hesse, Mark Chen, Eric Sigler, Mateusz Litwin, Scott Gray, Benjamin Chess, Jack Clark, Christopher Berner, Sam McCandlish, Alec Radford, Ilya Sutskever, and Dario Amodei.
\newblock Language models are few-shot learners.
\newblock In {\em Advances in Neural Information Processing Systems}, volume~33, pages 1877--1901, 2020.

\bibitem{shi2019}
Chuan Shi, Binbin Hu, Wayne~Xin Zhao, and Philip~S. Yu.
\newblock Heterogeneous information network embedding for recommendation.
\newblock {\em IEEE Transactions on Knowledge and Data Engineering}, 31:357--370, 2019.

\bibitem{zhang2016}
Fuzheng Zhang, Nicholas~Jing Yuan, Defu Lian, Xing Xie, and Wei-Ying Ma.
\newblock Collaborative knowledge base embedding for recommender systems.
\newblock In {\em Proceedings of the 22nd ACM SIGKDD International Conference on Knowledge Discovery and Data Mining}, pages 353--362, 2016.

\bibitem{wang2011}
Chong Wang and David~M. Blei.
\newblock Collaborative topic modeling for recommending scientific articles.
\newblock In {\em Proceedings of the 17th ACM SIGKDD International Conference on Knowledge Discovery and Data Mining}, pages 448--456, 2011.

\bibitem{freeman1977}
Linton~C. Freeman.
\newblock A set of measures of centrality based on betweenness.
\newblock {\em Sociometry}, pages 35--41, 1977.

\bibitem{yan2009}
Erjia Yan and Ying Ding.
\newblock Applying centrality measures to impact analysis: A coauthorship network analysis.
\newblock {\em Journal of the American Society for Information Science and Technology}, 60:2107--2118, 2009.

\bibitem{palshikar2007}
Girish~Keshav Palshikar.
\newblock Keyword extraction from a single document using centrality measures.
\newblock In {\em Pattern Recognition and Machine Intelligence}, pages 503--510, 2007.

\bibitem{bader2006}
David~A. Bader and Kamesh Madduri.
\newblock Parallel algorithms for evaluating centrality indices in real-world networks.
\newblock In {\em 2006 International Conference on Parallel Processing (ICPP'06)}, pages 539--550, 2006.

\bibitem{you2017}
Keyou You, Roberto Tempo, and Li~Qiu.
\newblock Distributed algorithms for computation of centrality measures in complex networks.
\newblock {\em IEEE Transactions on Automatic Control}, 62:2080--2094, 2017.

\bibitem{dedomenico2015}
Manlio De~Domenico, Vincenzo Nicosia, Alexandre Arenas, and Vito Latora.
\newblock Structural reducibility of multilayer networks.
\newblock {\em Nature Communications}, 6(1):6864, 2015.

\bibitem{stanley2016}
Natalie Stanley, Saray Shai, Dane Taylor, and Peter~J. Mucha.
\newblock Clustering network layers with the strata multilayer stochastic block model.
\newblock {\em IEEE Transactions on Network Science and Engineering}, 3:95--105, 2016.

\bibitem{kivela2014}
Mikko Kivel^^c3^^a4, Alex Arenas, Marc Barthelemy, James~P. Gleeson, Yamir Moreno, and Mason~A. Porter.
\newblock Multilayer networks.
\newblock {\em Journal of Complex Networks}, 2:203--271, 2014.

\bibitem{boccaletti2014}
S.~Boccaletti, G.~Bianconi, R.~Criado, C.I. {del Genio}, J.~G^^c3^^b3mez-Garde^^c3^^b1es, M.~Romance, I.~Sendi^^c3^^b1a-Nadal, Z.~Wang, and M.~Zanin.
\newblock The structure and dynamics of multilayer networks.
\newblock {\em Physics Reports}, 544:1--122, 2014.

\bibitem{iacopini2019}
Iacopo Iacopini, Giovanni Petri, Alain Barrat, and Vito Latora.
\newblock Simplicial models of social contagion.
\newblock {\em Nature Communications}, 10:2485, 2019.

\bibitem{dearruda2020}
Guilherme~Ferraz de~Arruda, Giovanni Petri, and Yamir Moreno.
\newblock Social contagion models on hypergraphs.
\newblock {\em Physical Review Research}, 2:023032, 2020.

\bibitem{landry2020}
Nicholas~W. Landry and Juan~G. Restrepo.
\newblock The effect of heterogeneity on hypergraph contagion models.
\newblock {\em Chaos: An Interdisciplinary Journal of Nonlinear Science}, 30:103117, 2020.

\bibitem{mancastroppa2023}
Marco Mancastroppa, Iacopo Iacopini, Giovanni Petri, and Alain Barrat.
\newblock Hyper-cores promote localization and efficient seeding in higher-order processes.
\newblock {\em Nature Communications}, 14:6223, 2023.

\bibitem{lucas2020}
Maxime Lucas, Giulia Cencetti, and Federico Battiston.
\newblock Multiorder laplacian for synchronization in higher-order networks.
\newblock {\em Physical Review Research}, 2:033410, 2020.

\bibitem{zhang2023}
Yuanzhao Zhang, Maxime Lucas, and Federico Battiston.
\newblock Higher-order interactions shape collective dynamics differently in hypergraphs and simplicial complexes.
\newblock {\em Nature Communications}, 14:1605, 2023.

\bibitem{xu2024}
Yan Xu, Dawei Zhao, Jiaxing Chen, Tao Liu, and Chengyi Xia.
\newblock The nested structures of higher-order interactions promote the cooperation in complex social networks.
\newblock {\em Chaos, Solitons \& Fractals}, 185:115174, 2024.

\bibitem{alvarezrodriguez2021}
Unai Alvarez-Rodriguez, Federico Battiston, Guilherme~Ferraz de~Arruda, Yamir Moreno, Matja{\v z} Perc, and Vito Latora.
\newblock Evolutionary dynamics of higher-order interactions in social networks.
\newblock {\em Nature Human Behaviour}, 5:586--595, 2021.

\bibitem{ruggeri2024}
Nicol\`o Ruggeri, Federico Battiston, and Caterina De~Bacco.
\newblock Framework to generate hypergraphs with community structure.
\newblock {\em Physical Review E}, 109:034309, 2024.

\bibitem{surana2023}
Amit Surana, Can Chen, and Indika Rajapakse.
\newblock Hypergraph similarity measures.
\newblock {\em IEEE Transactions on Network Science and Engineering}, 10:658--674, 2023.

\bibitem{lucas2024}
Maxime Lucas, Luca Gallo, Arsham Ghavasieh, Federico Battiston, and Manlio De~Domenico.
\newblock Functional reducibility of higher-order networks.
\newblock {\em arXiv preprint arXiv:2404.08547}, 2024.

\bibitem{feng2024}
Ruonan Feng, Tao Xu, Xiaowen Xie, Zi-Ke Zhang, Chuang Liu, and Xiu-Xiu Zhan.
\newblock A hyper-distance-based method for hypernetwork comparison.
\newblock {\em Chaos: An Interdisciplinary Journal of Nonlinear Science}, 34:083120, 2024.

\bibitem{agostinelli2025}
Cosimo Agostinelli, Marco Mancastroppa, and Alain Barrat.
\newblock Higher-order dissimilarity measures for hypergraph comparison.
\newblock {\em arXiv preprint arXiv:2503.16959}, 2025.

\bibitem{neuhauser2024}
Leonie Neuh^^c3^^a4user, Michael Scholkemper, Francesco Tudisco, and Michael~T. Schaub.
\newblock Learning the effective order of a hypergraph dynamical system.
\newblock {\em Science Advances}, 10:eadh4053, 2024.

\bibitem{kaminski2019}
Bogumi^^c5^^82 Kami^^c5^^84ski, Val^^c3^^a9rie Poulin, Pawe^^c5^^82 Pra^^c5^^82at, Przemys^^c5^^82aw Szufel, and Fran^^c3^^a7ois Th^^c3^^a9berge.
\newblock Clustering via hypergraph modularity.
\newblock {\em PLOS ONE}, 14:e0224307, 2019.

\bibitem{kaminski2024}
Bogumi^^c5^^82 Kami^^c5^^84ski, Pawe^^c5^^82 Misiorek, Pawe^^c5^^82 Pra^^c5^^82at, and Fran^^c3^^a7ois Th^^c3^^a9berge.
\newblock Modularity based community detection in hypergraphs.
\newblock {\em Journal of Complex Networks}, 12:cnae041, 2024.

\bibitem{latouche2014}
Pierre Latouche, Etienne Birmel{\'e}, and Christophe Ambroise.
\newblock {Model selection in overlapping stochastic block models}.
\newblock {\em Electronic Journal of Statistics}, 8:762--794, 2014.

\bibitem{peixoto2019}
Tiago~P. Peixoto.
\newblock {\em Bayesian Stochastic Blockmodeling}, chapter~11, pages 289--332.
\newblock 2019.

\bibitem{nakajima2022}
Kazuki Nakajima, Kazuyuki Shudo, and Naoki Masuda.
\newblock Randomizing hypergraphs preserving degree correlation and local clustering.
\newblock {\em IEEE Transactions on Network Science and Engineering}, 9:1139--1153, 2022.

\bibitem{lee2021}
Geon Lee, Minyoung Choe, and Kijung Shin.
\newblock How do hyperedges overlap in real-world hypergraphs? - {P}atterns, measures, and generators.
\newblock In {\em Proceedings of the Web Conference 2021}, pages 3396--3407, 2021.

\bibitem{malizia2025}
Federico Malizia, Santiago Lamata-Ot{\'\i}n, Mattia Frasca, Vito Latora, and Jes{\'u}s G{\'o}mez-Garde{\~n}es.
\newblock Hyperedge overlap drives explosive transitions in systems with higher-order interactions.
\newblock {\em Nature Communications}, 16:555, 2025.

\bibitem{nortier2025}
Bern{\'e}~L Nortier, Simon Dobson, and Federico Battiston.
\newblock Higher-order shortest paths in hypergraphs.
\newblock {\em arXiv preprint arXiv:2502.03020}, 2025.

\bibitem{gallo2024}
Luca Gallo, Lucas Lacasa, Vito Latora, and Federico Battiston.
\newblock Higher-order correlations reveal complex memory in temporal hypergraphs.
\newblock {\em Nature Communications}, 15:4754, 2024.

\bibitem{newman2016}
M.~E.~J. Newman and Aaron Clauset.
\newblock Structure and inference in annotated networks.
\newblock {\em Nature Communications}, 7:11863, 2016.

\bibitem{badalyan2024}
Anna Badalyan, Nicol{\`o} Ruggeri, and Caterina De~Bacco.
\newblock Structure and inference in hypergraphs with node attributes.
\newblock {\em Nature Communications}, 15:7073, 2024.

\bibitem{nakajima2025}
Kazuki Nakajima and Takeaki Uno.
\newblock Inference and visualization of community structure in attributed hypergraphs using mixed-membership stochastic block models.
\newblock {\em Social Network Analysis and Mining}, 15:5, 2025.

\bibitem{dempster1977}
A.~P. Dempster, N.~M. Laird, and D.~B. Rubin.
\newblock Maximum likelihood from incomplete data via the em algorithm.
\newblock {\em Journal of the Royal Statistical Society: Series B}, 39:1--22, 1977.

\bibitem{libennowell2007}
David Liben-Nowell and Jon Kleinberg.
\newblock The link-prediction problem for social networks.
\newblock {\em Journal of the American Society for Information Science and Technology}, 58:1019--1031, 2007.

\bibitem{chen2024}
Can Chen and Yang-Yu Liu.
\newblock A survey on hyperlink prediction.
\newblock {\em IEEE Transactions on Neural Networks and Learning Systems}, 35:15034--15050, 2024.

\bibitem{graham1989}
Ronald~L. Graham, Donald~E. Knuth, and Oren Patashnik.
\newblock {\em {Concrete mathematics: A foundation for computer science}}.
\newblock Addison-Wesley Longman Publishing Co., Inc., 1989.

\bibitem{acm_ccs}
{Association for Computing Machinery}.
\newblock {ACM Computing Classification System}, 2012.
\newblock \url{https://dl.acm.org/ccs}.

\bibitem{edge2024}
Darren Edge, Ha~Trinh, Newman Cheng, Joshua Bradley, Alex Chao, Apurva Mody, Steven Truitt, Dasha Metropolitansky, Robert~Osazuwa Ness, and Jonathan Larson.
\newblock From local to global: A graph rag approach to query-focused summarization.
\newblock {\em arXiv preprint arXiv:2404.16130}, 2024.

\end{thebibliography}

\begin{thebibliography}{1}

\bibitem{debacco2017}
Caterina De~Bacco, Eleanor~A. Power, Daniel~B. Larremore, and Cristopher Moore.
\newblock Community detection, link prediction, and layer interdependence in multilayer networks.
\newblock {\em Physical Review E}, 95:042317, 2017.

\bibitem{ruggeri2023}
Nicol^^c3^^b2 Ruggeri, Martina Contisciani, Federico Battiston, and Caterina~De Bacco.
\newblock Community detection in large hypergraphs.
\newblock {\em Science Advances}, 9:eadg9159, 2023.

\bibitem{newman2016}
M.~E.~J. Newman and Aaron Clauset.
\newblock Structure and inference in annotated networks.
\newblock {\em Nature Communications}, 7:11863, 2016.

\bibitem{contisciani2022}
Martina Contisciani, Federico Battiston, and Caterina De~Bacco.
\newblock Inference of hyperedges and overlapping communities in hypergraphs.
\newblock {\em Nature Communications}, 13:7229, 2022.

\bibitem{nakajima2025}
Kazuki Nakajima and Takeaki Uno.
\newblock Inference and visualization of community structure in attributed hypergraphs using mixed-membership stochastic block models.
\newblock {\em Social Network Analysis and Mining}, 15:5, 2025.

\bibitem{lotito2023}
Quintino~Francesco Lotito, Martina Contisciani, Caterina De~Bacco, Leonardo Di~Gaetano, Luca Gallo, Alberto Montresor, Federico Musciotto, Nicol^^c3^^b2 Ruggeri, and Federico Battiston.
\newblock Hypergraphx: {A} library for higher-order network analysis.
\newblock {\em Journal of Complex Networks}, 11:cnad019, 2023.

\end{thebibliography}
\end{document}